\documentclass[sigconf]{acmart}

\setcopyright{acmcopyright}
\copyrightyear{2026}
\acmYear{2026}
\acmDOI{10.1145/XXXXXX.XXXXXX}  
\acmConference[ICSE '26]{2026 IEEE/ACM 48th International Conference on Software Engineering}{April 12--18, 2026}{Rio de Janeiro, Brazil}
\acmBooktitle{2026 IEEE/ACM 48th International Conference on Software Engineering (ICSE '26), April 12--18, 2026, Rio de Janeiro, Brazil}
\acmPrice{15.00} 
\acmISBN{979-X-XXXX-XXXX-X/26/04}

\usepackage{algpseudocode}
\usepackage{amsmath}  
\usepackage[ruled,linesnumbered]{algorithm2e}
\usepackage{multirow}
\usepackage{algpseudocode}
\usepackage{tikz}
\newcommand{\circnumber}[1]{\lower.75ex\hbox{\tikz\draw (0pt, 0pt)%
    circle (.47em) node {\makebox[.15em][c]{\small #1}};}}
\newcommand{\ballnumber}[1]{\lower.75ex\hbox{\tikz\fill(0pt, 0pt)%
    circle (.5em) node {\makebox[.15em][c]{\small \textcolor{white}{#1}}};}}
\newcommand{\dcircnumber}[1]{\lower.75ex\hbox{\tikz\draw(0pt, 0pt)%
    circle (.47em) circle (.37em) node {\makebox[.15em][c]{\small #1}};}}

\newcommand{\emptycirc}[1]{\lower.75ex\hbox{\tikz\draw (0pt, 0pt)%
    circle (.47em) node {\makebox[.15em][c]{\small \textcolor{white}{#1}}};}}
\newcommand{\emptyball}[1]{\lower.75ex\hbox{\tikz\fill(0pt, 0pt)%
    circle (.5em) node {\makebox[.15em][c]{\small #1}};}}
\newcommand{\emptydcirc}[1]{\lower.75ex\hbox{\tikz\draw(0pt, 0pt)%
    circle (.47em) circle (.37em) node {\makebox[.15em][c]{\small \textcolor{white}{#1}}};}}
            
\usepackage{listings, xcolor}
\definecolor{verylightgray}{rgb}{.97,.97,.97}
\lstdefinelanguage{Solidity}{
  keywords=[1]{anonymous, assembly, assert, balance, break, call, callcode, case, catch, class, constant, continue, constructor, contract, debugger, default, delegatecall, delete, do, else, emit, event, experimental, export, external, false, finally, for, function, gas, if, implements, import, in, indexed, instanceof, interface, internal, is, length, library, log0, log1, log2, log3, log4, memory, modifier, new, payable, pragma, private, protected, public, pure, push, require, return, returns, revert, selfdestruct, send, solidity, storage, struct, suicide, super, switch, then, this, throw, transfer, true, try, typeof, using, value, view, while, with, addmod, ecrecover, keccak256, mulmod, ripemd160, sha256, sha3}, % generic keywords including crypto operations
  keywordstyle=[1]\color{blue}\bfseries,
  keywords=[2]{address, bool, byte, bytes, bytes1, bytes2, bytes3, bytes4, bytes5, bytes6, bytes7, bytes8, bytes9, bytes10, bytes11, bytes12, bytes13, bytes14, bytes15, bytes16, bytes17, bytes18, bytes19, bytes20, bytes21, bytes22, bytes23, bytes24, bytes25, bytes26, bytes27, bytes28, bytes29, bytes30, bytes31, bytes32, enum, int, int8, int16, int24, int32, int40, int48, int56, int64, int72, int80, int88, int96, int104, int112, int120, int128, int136, int144, int152, int160, int168, int176, int184, int192, int200, int208, int216, int224, int232, int240, int248, int256, mapping, string, uint, uint8, uint16, uint24, uint32, uint40, uint48, uint56, uint64, uint72, uint80, uint88, uint96, uint104, uint112, uint120, uint128, uint136, uint144, uint152, uint160, uint168, uint176, uint184, uint192, uint200, uint208, uint216, uint224, uint232, uint240, uint248, uint256, var, void, ether, finney, szabo, wei, days, hours, minutes, seconds, weeks, years},  % types; money and time units
  keywordstyle=[2]\color{teal}\bfseries,
  keywords=[3]{block, blockhash, coinbase, difficulty, gaslimit, number, timestamp, msg, data, gas, sender, sig, value, now, tx, gasprice, origin},  % environment variables
  keywordstyle=[3]\color{violet}\bfseries,
  identifierstyle=\color{black},
  sensitive=false,
  comment=[l]{//},
  morecomment=[s]{/*}{*/},
  commentstyle=\color{gray}\ttfamily,
  stringstyle=\color{red}\ttfamily,
  morestring=[b]',
  morestring=[b]"
}
\usepackage{breakurl}%此行一定要在hyperref包之后！
\usepackage{pifont}

\AtBeginDocument{%
  }

\begin{document}

\title{\textit{One Signature, Multiple Payments:} Demystifying and Detecting Signature Replay Vulnerabilities in Smart Contracts}

\author{Zexu Wang}
\orcid{0009-0004-1439-2989}
\affiliation{%
  \institution{Sun Yat-sen University}
  \city{Zhuhai}
  \country{China}
}
\affiliation{%
  \institution{Peng Cheng Laboratory}
  \city{Shenzhen}
  \country{China}
}
\email{wangzx97@mail2.sysu.edu.cn}

\author{Jiachi Chen}
\orcid{0000-0002-0192-9992}
\affiliation{%
  \institution{Sun Yat-sen University}
  \city{Zhuhai}
  \country{China}
}
\affiliation{%
  \institution{Zhejiang University}
  \city{Hangzhou}
  \country{China}
}
\email{chenjch86@mail.sysu.edu.cn}

\author{Zewei Lin}
\orcid{0009-0008-4095-5772}
\affiliation{%
  \institution{Sun Yat-sen University}
  \city{Zhuhai}
  \country{China}
}
\affiliation{%
  \institution{Peng Cheng Laboratory}
  \city{Shenzhen}
  \country{China}
}
\email{linzw3@mail2.sysu.edu.cn}

\author{Wenqing Chen}
\orcid{0000-0002-8739-2216}
\authornote{Corresponding Author}
\affiliation{%
  \institution{Sun Yat-sen University}
  \city{Zhuhai}
  \country{China}
}
\email{chenwq95@mail.sysu.edu.cn}

\author{Kaiwen Ning}
\orcid{0009-0009-6009-8285}
\affiliation{%
  \institution{Sun Yat-sen University}
  \city{Zhuhai}
  \country{China}
}
\affiliation{%
  \institution{Peng Cheng Laboratory}
  \city{Shenzhen}
  \country{China}
}
\email{ningkw@mail2.sysu.edu.cn}

\author{Jianxing Yu}
\orcid{0000-0003-1340-3995}
\affiliation{%
  \institution{Sun Yat-sen University}
  \city{Zhuhai}
  \country{China}
}
\email{yujx26@mail.sysu.edu.cn}

\author{Yuming Feng}
\orcid{0000-0001-8922-0496}
\affiliation{%
  \institution{Peng Cheng Laboratory}
  \city{Shenzhen}
  \country{China}
}
\email{fengym@pcl.ac.cn}

\author{Yu Zhang}
\orcid{0000-0003-2040-5059}
\affiliation{%
  \institution{Harbin Institute of Technology}
  \city{Harbin}
  \country{China}
}
\affiliation{%
  \institution{Peng Cheng Laboratory}
  \city{Shenzhen}
  \country{China}
}
\email{yuzhang@hit.edu.cn}

\author{Weizhe Zhang}
\orcid{0000-0003-4783-876X}
\affiliation{%
  \institution{Harbin Institute of Technology}
  \city{Harbin}
  \country{China}
}
\affiliation{%
  \institution{Peng Cheng Laboratory}
  \city{Shenzhen}
  \country{China}
}
\email{wzzhang@hit.edu.cn}

\author{Zibin Zheng}
\orcid{0000-0002-7878-4330}
\affiliation{%
  \institution{Sun Yat-sen University, Guangdong Engineering Technology Research Center of Blockchain}
  \city{Zhuhai}
  \country{China}
}
% \affiliation{%
%   \institution{Guangdong Engineering Technology Research Center of Blockchain}
%   \city{Zhuhai}
%   \country{China}
% }
\email{zhzibin@mail.sysu.edu.cn}

\renewcommand{\shortauthors}{Wang et al.}

\begin{abstract}
  Smart contracts have significantly advanced blockchain technology, and digital signatures are crucial for reliable verification of contract authority. Through signature verification, smart contracts can ensure that signers possess the required permissions, thus enhancing security and scalability. However, lacking checks on signature usage conditions can lead to repeated verifications, increasing the risk of permission abuse and threatening contract assets. We define this issue as the \textit{Signature Replay Vulnerability} (SRV).
  
  In this paper, we conducted the first empirical study to investigate the causes and characteristics of the SRVs. From 1,419 audit reports across 37 blockchain security companies, we identified 108 with detailed SRV descriptions and classified five types of SRVs. To detect these vulnerabilities automatically, we designed \textit{LASiR}, which utilizes the general semantic understanding ability of \textit{Large Language Models} (LLMs) to assist in the static taint analysis of the signature state and identify the signature reuse behavior. It also employs path reachability verification via symbolic execution to ensure effective and reliable detection. To evaluate the performance of \textit{LASiR}, we conducted large-scale experiments on 15,383 contracts involving signature verification, selected from the initial dataset of 918,964 contracts across four blockchains: \textit{Ethereum}, \textit{Binance Smart Chain}, \textit{Polygon}, and \textit{Arbitrum}. The results indicate that SRVs are widespread, with affected contracts holding \$4.76 million in active assets. Among these, 19.63\% of contracts that use signatures on \textit{Ethereum} contain SRVs. Furthermore, manual verification demonstrates that \textit{LASiR} achieves an \textit{F1-score} of 87.90\% for detection. Ablation studies and comparative experiments reveal that the semantic information provided by LLMs aids static taint analysis, significantly enhancing \textit{LASiR}'s detection performance.

\end{abstract}



\maketitle

\section{Introduction}
% As the crucial program for blockchain state calculation, smart contract security directly impacts numerous on-chain assets. 
% Digital signatures provide secure verification of transactions and data through encryption and hashing, addressing challenges like identity authentication, tamper-proofing, and data protection. Furthermore, they enable innovative DeFi designs—such as multi-signature wallets~\cite{multisig}, cross-chain interactions~\cite{crosschain}, and ERC20 with permits~\cite{erc20permit}—enhancing user experience, security, and flexibility. 

Smart contracts, as programs running on the blockchain, directly impact the security of digital assets. Digital signatures are commonly used in contracts to verify that the signer has the required permissions. They employ encryption and hashing to securely verify transactions and data, enhancing both security and scalability. Digital signatures have been widely used in contract development for multi-signatures in asset management~\cite{multisig}, maintaining information integrity between blockchains~\cite{crosschain}, and enabling functions like "checks" in digital assets~\cite{erc20permit}, among others.

% While digital signatures are widely used in smart contracts, signature replay attacks have become increasingly severe due to flaws in improper signature verification implementation. 
While digital signatures are widely used, the lack of checks on signature usage conditions (such as user identity requirements and the validity period) compromises the uniqueness of signature verification. This can result in the verification of a single signature multiple times, leading to \textit{Signature Replay Vulnerabilities} (SRVs). In such scenarios, attackers reuse the same signature to pass multiple authorization checks for payments, illegally obtaining assets and compromising user trust and assets. A significant factor contributing to these vulnerabilities is developers' insufficient understanding of signature security practices. The research by Zhang et al.~\cite{cryptoapi} reveals that 56.3\% of contract developers face struggles to implement cryptographic practices, and 68.1\% believe that the existing security tools need improvement. Many real-world developers are not cryptographic experts and lack experience in security development, leading to frequent signature replay attacks. For example, in July 2023, the \textit{AzukiDAO} project lacked checks on signature usage, resulting in reused signatures and the \$69K asset loss~\cite{azukidao}. Additionally, the signature verification status is often dispersed throughout the codebase and requires semantic analysis for validation, which attackers can exploit if there are errors or omissions in the verification process. For example, due to a developer error, the \textit{branchMask} function in the \textit{Polygon Plasma Bridge} generated the same signature for different \textit{branch masks}, allowing the signature to be validated 223 times for burn transactions. This malleability of the signature enabled an attacker to steal \$22.3 million~\cite{polygondoublespend}.

% For example, the \textit{branchMask} function in the \textit{Polygon Plasma Bridge} erroneously encodes different branch masks as the same \textit{Exit ID} signature, allowing 223 valid \textit{Exit ID} signatures for a single burn transaction. This malleability of the signature allowed an attacker to steal \$22.3 million~\cite{polygondoublespend}.

% For example, in July 2022, the Quixotic NFT Marketplace project on the Optimism blockchain was hacked, leading to the theft of approximately \$100K worth of Bean tokens due to incorrect seller identity checks in signature verification~\cite{quixotic}.

% Signature replay vulnerabilities present new challenges for smart contract security. First, there is a lack of systematic research on these vulnerabilities, making it difficult to summarize characteristics and design detection rules. Second, signatures are required across various scenarios, and verification cannot be repeated. The complex signature information must align with the execution context. Third, analyzing dangerous signature verification patterns in complex contracts requires a comprehensive understanding of program semantics and contract intent, which is highly challenging.

\textit{Signature Replay Vulnerabilities} present new challenges for contract security. First, there is a lack of systematic research, making it difficult to summarize characteristics and design detection rules. Second, analyzing dangerous signature verification patterns in complex contracts requires a comprehensive understanding of program semantics and contract intent, which is highly challenging.

% To address these challenges, we empirically studied open-source contract audit reports from security companies, summarizing and classifying dangerous patterns that lead to signature reuse. Specifically, we manually examined 1,419 public audit reports, identifying 256 reports related to signature verification and analyzing them using the Card Sorting method. We categorized five types of signature replay vulnerabilities: \textit{Cross-chain Replay Attack} (X-CRA), \textit{Cross-project Replay Attack} (X-PRA), \textit{Contract Account Signature Replay} (CASR), \textit{Signature State Management Issue} (SSMI), and \textit{Signature Malleability Attack} (SMA) (see Section~\ref{subsec:define} for details). These vulnerability types, derived from real audit reports, represent issues encountered by developers in the real world.

% To address these challenges, we manually examined 1,419 open-source contract audit reports from security companies, analyzing 256 reports related to signature verification using the \textit{Card Sorting} method~\cite{cardsorting} to classify dangerous patterns leading to signature reuse. We categorized five types of SRVs: \textit{Cross-chain Replay Attack} (X-CRA), \textit{Cross-project Replay Attack} (X-PRA), \textit{Contract Account Signature Replay} (CASR), \textit{Signature State Management Issue} (SSMI), and \textit{Signature Malleability Attack} (SMA) (see Subsection~\ref{subsec:define} for details).

To address these challenges, we conducted the first empirical study to summarize the causes and definitions of SRVs. Reflecting real developer issues, we manually examined 1,419 open-source contract audit reports from 37 security companies and identified 108 reports related to the reuse of signatures. Using the Open Card Sorting method~\cite{cardsorting}, we classified dangerous patterns leading to signature reuse issues during verification. Finally, we identified five types of SRVs: \textit{Cross-chain Replay Attack} (X-CRA), \textit{Cross-project Replay Attack} (X-PRA), \textit{Contract Account Signature Replay} (CASR), \textit{Signature State Management Issue} (SSMI), and \textit{Signature Malleability Attack} (SMA) (see Subsection~\ref{subsec:define} for details).

To effectively detect SRVs, we designed a tool named \textit{LASiR}, which utilizes \textit{Large Language Models} (LLMs)~\cite{llmintro} to understand contract semantics, combining static taint analysis and symbolic execution to enhance detection reliability. It inputs smart contract source code and outputs detection results in three phases: \textit{Slicing with LLM Analysis}, \textit{Inspection of Signature Verification}, and \textit{Path Reachability Verification}. In \textit{Phase 1}, \textit{LASiR} utilizes LLMs to identify variables related to signature states and analyzes their dependencies to perform program slicing. In \textit{Phase 2}, \textit{LASiR} leverages LLMs to analyze the semantic information within the slices related to signature verification. It identifies sanitized variables to assist in taint analysis status checks, detects hazardous signature verification patterns, and generates \textit{Warnings}. In \textit{Phase 3}, \textit{LASiR} requires LLMs to review \textit{Warnings} and provide function sequences to guide symbolic execution for path reachability verification. \textit{LASiR} leverages LLMs' general understanding ability to assist in static taint analysis, ensuring detection accuracy and reliability.

We conducted three experiments to evaluate \textit{LASiR}'s detection performance for SRVs. First, we crawled 918,964 contract source codes from \textit{Ethereum}~\cite{ethereum}, \textit{Binance Smart Chain}~\cite{bsc}, \textit{Polygon}~\cite{polygon}, and \textit{Arbitrum}~\cite{arbitrum} to analyze performance on large-scale datasets. 
% By examining the use of the \textit{ecrecover()} function in the contract's AST file, 
By analyzing the contract's AST file, we screened 15,383 contracts related to signature verification (\textit{DB1}). Experiments revealed that SRVs are widespread, with affected contracts holding \$4.76 million in active assets. Among these, 19.63\% of contracts that use signatures on \textit{Ethereum} contain SRVs. The average detection time is approximately 40 seconds, with a total LLM API cost of around \$15, demonstrating \textit{LASiR}'s efficiency and low cost for detection. To further analyze \textit{LASiR}'s effectiveness, we randomly selected 500 contracts from \textit{DB1} for manual analysis, identifying 72 positive and 428 negative cases (\textit{DB2}). \textit{LASiR} achieved a \textit{Precision} of 82.14\%, a \textit{Recall} of 95.83\%, and an \textit{F1-score} of 88.46\%, outperforming the compared general-purpose tools. Additionally, ablation experiments analyzing the impact of LLM on \textit{LASiR}'s performance showed significant improvements: \textit{Precision} increased from 4.40\% to 82.14\%, \textit{Recall} from 26.39\% to 95.83\%, and \textit{F1-score} from 7.54\% to 88.46\%. The information provided by LLM through contract context analysis is essential for static taint analysis, enhancing accuracy and efficiency.

The main contributions of this work are as follows:
\vspace{-0.3cm}
\begin{itemize}
    \item This study conducted the first empirical analysis of SRVs in smart contracts. We manually examine real-world security audit reports, define five types of SRVs, and provide explanatory examples.
    \item We designed \textit{LASiR}, leveraging LLM's semantic understanding to assist in static taint analysis of the signature state, achieving efficient detection of SRVs.
    \item We provide a dataset of real-world SRVs from 918,964 contracts across four blockchains. This dataset identifies 1,739 contracts with SRVs holding \$4.76 million in assets, which can further aid research in vulnerability repair and automated exploitation efforts.
    \item We have open-source \textit{LASiR}'s tool code, experimental data, and empirical research data at \url{https://anonymous.4open.science/r/LASiR-B207}.
\end{itemize}

% \vspace{-10pt}
\section{Background}
\subsection{Signature Verification of Smart Contracts}
% ECDSA (Elliptic Curve Digital Signature Algorithm)~\cite{ecdsa} is a widely used encryption algorithm in smart contracts. It enables a signer to use the private key to sign a message, while anyone can verify the signature using the signer's public key, ensuring transaction authenticity and integrity. 
The digital signature is an effective method to verify the identity of the signer~\cite{wiki:DigitalSignature}. For example, contracts using the \textit{Elliptic Curve Digital Signature Algorithm} (ECDSA)~\cite{ecdsa} allow the signer to sign a message with the private key, while anyone can verify the signature with the public key, ensuring the authenticity and integrity of the signed message. \textit{Ethereum} deploys the verifying signature program in \textit{ecrecover()} (precompiled contract at address \textit{0x1}) for on-chain verification~\cite{ecrecoverfunction}. \textit{ecrecover()} inputs the message hash (\textit{\_messageHash}) and signature parameters (v, r, s\textit{}), recovers the signer's address. The message hash (\textit{\_messageHash}) is derived from the \textit{Keccak-256} hash~\cite{keccak256} operation on the signed message. The signature parameters (\textit{v, r, s}) result from the signer signing the \textit{\_messageHash}, following the \textit{Secp256k1} cryptographic algorithms~\cite{secp256k1}. Developers can directly call \textit{ecrecover()} in contracts to recover the signer's address, enabling identity verification and token delegation authorization (similar to "checks" in digital assets)~\cite{delegatecoin}, among others. 

\vspace{-0.3cm}
\begin{figure}[htb]
    \centering
    \includegraphics[width=\linewidth]{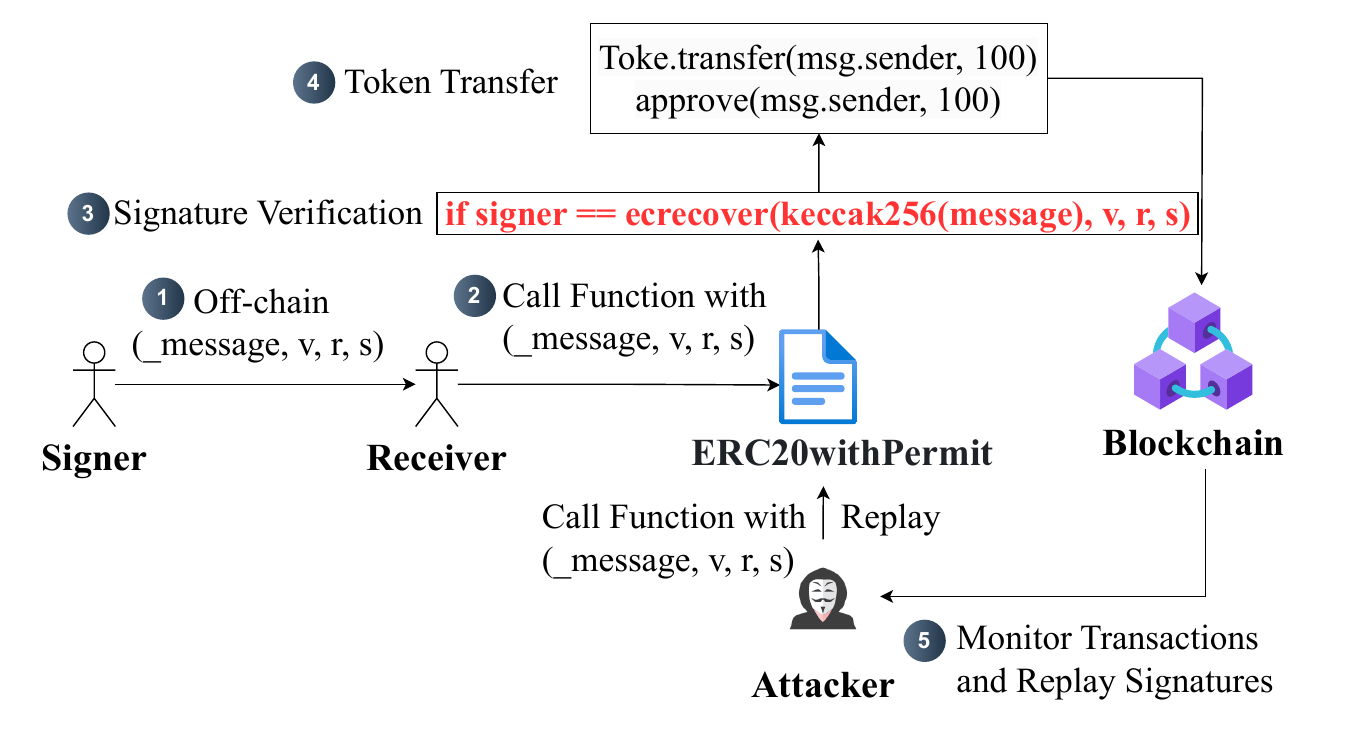}
    \caption{Signature Replay Attack Process.}
    \label{fig:bgexample}
\end{figure}
\vspace{-0.3cm}

Figure~\ref{fig:bgexample} illustrates the signature replay attack process. The \textit{ERC20withPermit} contract, derived from the ERC-20 token standard~\cite{erc20}, allows users to authorize other accounts to transfer tokens without additional transactions. The process is outlined as follows: \ding{182} The signer signs the \textit{\_message}, generates the signature information \textit{(v, r, s)}, and notifies the receiver through an off-chain channel. \ding{183} The receiver submits the message (\textit{\_message}) and signature information (\textit{v, r, s}) as input parameters to the \textit{ERC20withPermit} contract, requesting a token transfer. \ding{184} The contract uses the \textit{ecrecover()} function to recover the address from the signature and message, checking if the recovered address matches the signer's address. \ding{185} If the verification is successful, the contract authorizes the \textit{msg.sender} (receiver) to execute the transfer operation with a limit of 100 tokens. However, signature verification lacks identity checks on the \textit{msg.sender}, allowing anyone to reuse the signature and posing significant threats to asset security. As shown in Figure~\ref{fig:bgexample}, \ding{186} an attacker can monitor blockchain transactions to obtain the message (\textit{\_message}) and signature information (\textit{v, r, s}) and reuse them to submit a transfer request to the \textit{ERC20withPermit} contract. Due to the lack of identity checks on the \textit{msg.sender}, the contract permits the attacker (\textit{msg.sender}) to transfer 100 tokens. 

% To prevent signature replay, it is crucial to check the usage conditions of the signature to ensure the uniqueness of the verification process. Integrating the user identity, validity period, usage status, and transaction details (such as amount and identities) into the signed message and validating this information during signature verification prevents potential signature reuse.

% \subsection{Static Taint Analysis \& LLMs}
\subsection{Static Taint Analysis}
Static taint analysis tracks data flow through predefined patterns to detect security vulnerabilities. The process comprises three main steps: identifying taint sources, analyzing taint propagation, and identifying and checking sinks. In smart contracts, taint sources include user inputs (e.g., \textit{msg.value}, \textit{msg.data}, \textit{msg.sender}), external blockchain attributes (e.g., \textit{block.timestamp}, \textit{block.chainid}), and return values from external contract calls (e.g., \textit{call}, \textit{delegatecall}). Sinks need to be custom-defined based on specific detection tasks and expert rules. Taint propagation analysis involves data flow and control flow analysis. Data flow analysis tracks the flow of tainted data through assignments and function calls, while control flow analysis examines whether tainted data reaches sensitive operations. This method effectively analyzes dependencies within smart contracts, making it a robust approach for detecting vulnerabilities. However, the accuracy of detection is highly dependent on the precise extraction of contract semantics. Many methods use fixed pattern matching for semantic feature extraction, resulting in challenges such as low automation and insufficient semantic analysis capabilities. Patterns heavily rely on manual experience, further constraining detection capabilities in complex scenarios.

\section{Signature Replay Vulnerability Definition}
In this section, we conduct an empirical study on real-world security audit reports regarding signature reuse to define and classify common \textit{Signature Replay Vulnerabilities} (SRVs) in smart contracts.

\subsection{Data Collection}
To comprehensively analyze real-world issues related to signature reuse, we collected open-source security audit reports from various security teams. Specifically, we accessed the public URLs (official websites, X (Twitter), and GitHub) of 81 smart contract security teams listed by Etherscan~\cite{etherscan}. Of these, 37 teams had public audit reports, including \textit{BlockSec}~\cite{blocksec}, \textit{Trail of Bits}~\cite{trailofbits}, and \textit{SlowMist}~\cite{slowmist}. Additionally, we gathered vulnerability audit reports from bug bounties publicly available on \textit{Solodit}~\cite{SoloditBugBounties}. In total, we manually collected 1,419 security audit reports.

\subsection{Data Pre-processing}

To filter security reports related to SRVs, we combined keyword filtering with manual checks. Initially, we selected commonly used terms in signature verification as keywords, including ``\textit{ecrecover()}", ``\textit{signature}", and ``\textit{replay attack}", with ``\textit{ecrecover()}" specifically chosen for its significance in signature verification. Automated keyword filtering identified 557 reports (467 with ``\textit{signature}", 28 with ``\textit{ecrecover()}", and 62 with ``\textit{replay attack}"), each containing at least one keyword. However, the multiple meanings of keywords can easily lead to misidentifications, as some reports may contain these keywords but are unrelated to signature replay. For example, the report~\cite{SoloditIncorrectFunctionSignature} highlights an issue with the "incorrect function signature", which was selected due to the ``\textit{signature}" keyword matching but is unrelated to SRVs. Therefore, manual checking is necessary to eliminate these irrelevant reports. Finally, through manual filtering, we obtained 108 security audit reports related to SRVs.

% We used keyword searches and manual checks to filter 256 security audit reports related to signature verification. Keywords included \textit{"ecrecover()”}, \textit{"signature”}, and \textit{"replay attack”}, specifically choosing \textit{"ecrecover()"} for its significance in signature verification. However, relying solely on keyword matching can easily select unrelated reports, so we manually reviewed and confirmed the reports to ensure that they were relevant to signature verification.

\subsection{Data Analysis}
\begin{figure}[ht]
    \centering
    \vspace{-6pt}
    \includegraphics[width=0.98\linewidth]{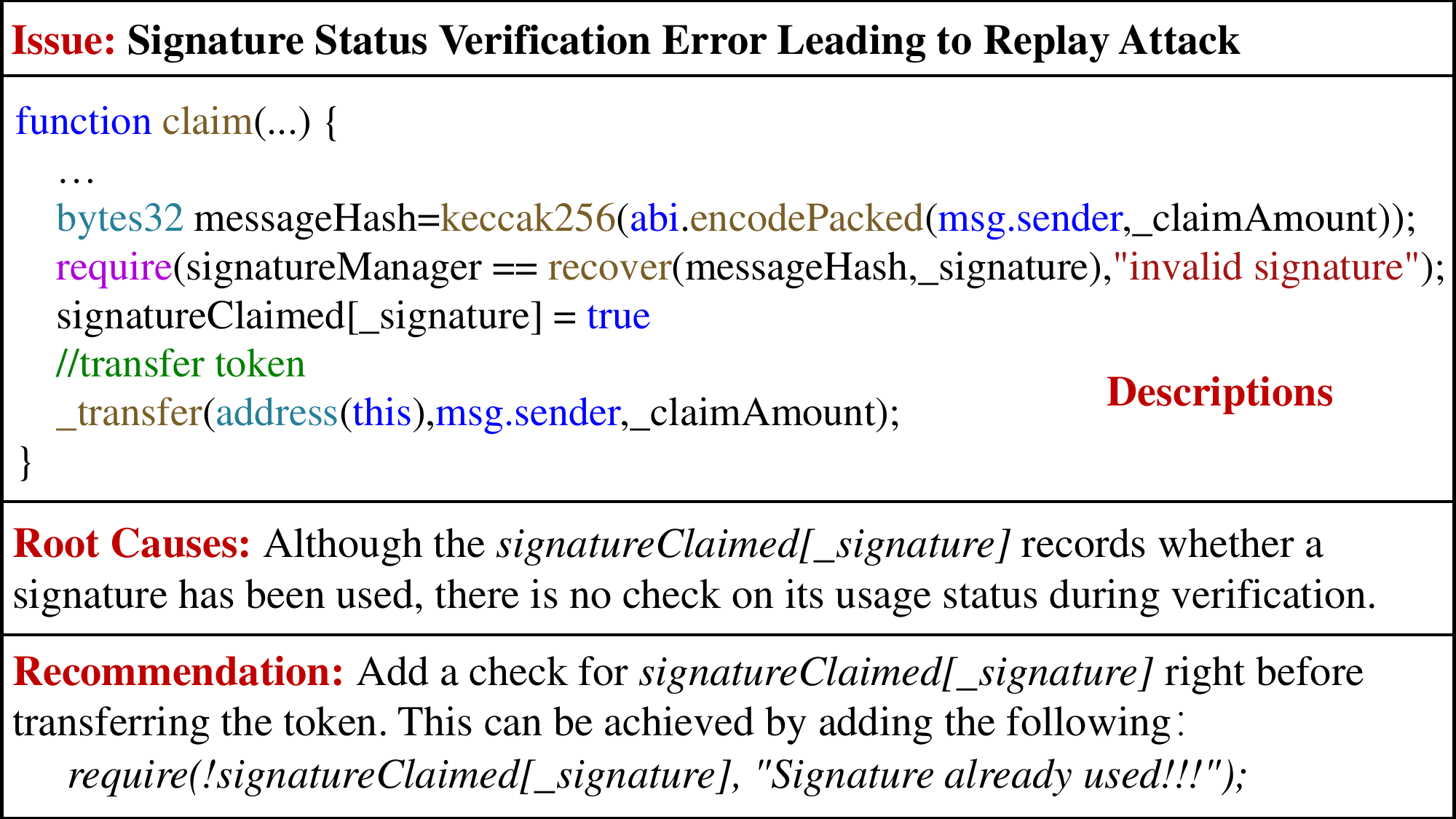}
    % \vspace{-1em}
    \caption{Example of the Audit Report Card.}
   % \vspace{-0.5em}
    \label{fig:cardexample}
\end{figure}

To classify SRVs, we used the Open Card Sorting method~\cite{cardsorting}, widely employed for problem discovery and definition in software engineering~\cite{chen2020defining,crpwarner,zhang2024demystifying}. We created a card for each audit report, detailing the \textit{Title}, \textit{Descriptions}, \textit{Root Causes}, and \textit{Recommendations}. For example, Figure~\ref{fig:cardexample} presents the card content of the report~\cite{cardexample}, highlighting key sections such as the \textit{Title} and \textit{Root Causes}, which directly indicate the lack of checks on the \textit{signatureClaimed[\_signature]} status, leading to signature reuse. The \textit{Descriptions} provides relevant case analysis, and the \textit{Recommendation} section outlines the mitigation measures. These professionally-audited reports offer structured content for quick vulnerability analysis.

% \vspace{-0.15cm}
\vspace{-0.15cm}

To accurately and objectively characterize real-world vulnerabilities, we adopt a classification framework grounded in three key dimensions: (i) commonly accepted industry terminology, (ii) prevalent developer challenges, and (iii) specific exploitation conditions. Domain experts with extensive experience in smart contract security ensure alignment with the language used in professional audit reports. For instance, \emph{signature malleability} is categorized based on its cryptographic underpinnings in the Elliptic Curve Digital Signature Algorithm (ECDSA). We further emphasize that various signature-related vulnerabilities (SRVs) originate from distinct root causes and triggering mechanisms. For example, \texttt{X-CRA} and \texttt{X-PRA} stem from separate misconfigurations in blockchain identity verification and project address validation, respectively. Moreover, we observe that vulnerabilities such as \emph{Signature State Management Issues} (SSMI) frequently arise due to developers' limited security awareness and flawed design logic, particularly in managing complex digital signature states within smart contracts. For a comprehensive explanation of the classification criteria and illustrative examples, please refer to Part~\uppercase\expandafter{\romannumeral1} of \textit{Appendix~A}~\cite{gitrepository}.

% \vspace{-0.15cm}
% \begin{table}[H]
%     % \setlength{\abovecaptionskip}{0.2cm}
%     % \setlength{\belowcaptionskip}{-0.2cm}
%     \centering
%     \caption{Statistical Distribution of Different SRVs Types.}
%     \vspace{-5pt}
%     \resizebox{0.88\linewidth}{!}{%
% \begin{tabular}{l|lllll}
% \hline
% \textbf{Type}     & \textbf{X-CRA} & \textbf{X-PRA} & \textbf{CASR} & \textbf{SSMI} & \textbf{SMA} \\ \hline
% \textbf{Quantity} & 33             & 1              & 48            & 5             & 21           \\ \hline
% \end{tabular}
%     }
%     \label{tab:SRV-distribution}
% \end{table}
% \vspace{-0.15cm}

% We invited two experienced contract security researchers to classify vulnerabilities without predefined categories. \ding{182} In the first round, they randomly selected 40\% of the cards for classification. They read the \textit{Title} and \textit{Descriptions} to understand the vulnerabilities, examined the problematic code to analyze the \textit{Root Causes}, and reviewed the \textit{Recommendations} to identify related types. If a vulnerability did not fit existing categories, they assessed its representativeness and repeatability before creating a new category. \ding{183} In the second round, they independently classified the remaining 60\% of the cards following the same steps as in the first round. \ding{184} They compared the labeling results of the cards, discussed discrepancies, and reached a consensus. Ultimately, they categorized them into five types. The labeling results are available in our repository~\cite{gitrepository}.

We invited two experienced smart contract security researchers to classify the vulnerabilities without predefined categories.

\ding{182} \textbf{First Round}. The researchers randomly selected 40\% of the vulnerability cards for initial classification. They examined each card by reading the \emph{Title} and \emph{Description} to understand the vulnerability, analyzing the problematic code to identify the \emph{Root Cause}, and reviewing the \emph{Recommendations} to infer related types. If a vulnerability did not fit any existing category, they evaluated its representativeness and recurrence before proposing a new category. This round resulted in five preliminary types of signature replay vulnerabilities (SRVs): \emph{Cross-Chain Replay Attacks}, \emph{Cross-Project Replay Attacks}, \emph{Signature State Management Issues}, \emph{Signature Malleability Attacks}, and \emph{Front-Running Replay Attacks}. Since all cards were sourced from real-world security audit reports with clearly defined content, no cards were excluded as irrelevant.

\ding{183} \textbf{Second Round.} The researchers independently classified the remaining 60\% of the cards. During this process, they identified an additional category: \emph{Contract Account Signature Replay}.

\ding{184} \textbf{Reconciliation.} The researchers then compared their labeling results. The main disagreements involved four cards, primarily concerning the distinction between \emph{Front-Running Replay} and \emph{Contract Account Signature Replay}. After discussion, they agreed that the latter exhibited distinct exploitation characteristics and had broad real-world impact, warranting its recognition as a standalone category. Conversely, the definition of \emph{Front-Running Replay} was deemed overly narrow, and its instances could be subsumed under other categories. As a result, this category was removed. Ultimately, five SRV types were finalized. 
% Table~\ref{tab:SRV-distribution} summarizes the distribution of these five SRV types across the audit reports. 
During the classification of all 108 cards, the two researchers disagreed on only 4 cards, yielding a disagreement rate of \(D = \frac{N_{\mathrm{disagree}}}{N_{\mathrm{total}}} = \frac{4}{108} \approx 3.7\%\), which reflects the clarity and consistency of the audit report content. The complete labeling results, along with the classification criteria and examples (detailed in Part~\uppercase\expandafter{\romannumeral2} of \textit{Appendix~A}), are available in repository~\cite{gitrepository}.

\subsection{Definition of Signature Replay Vulnerability}
\label{subsec:define}
\begin{figure}[htb]
    \centering
    \vspace{-6pt}
    \includegraphics[width=0.98\linewidth]{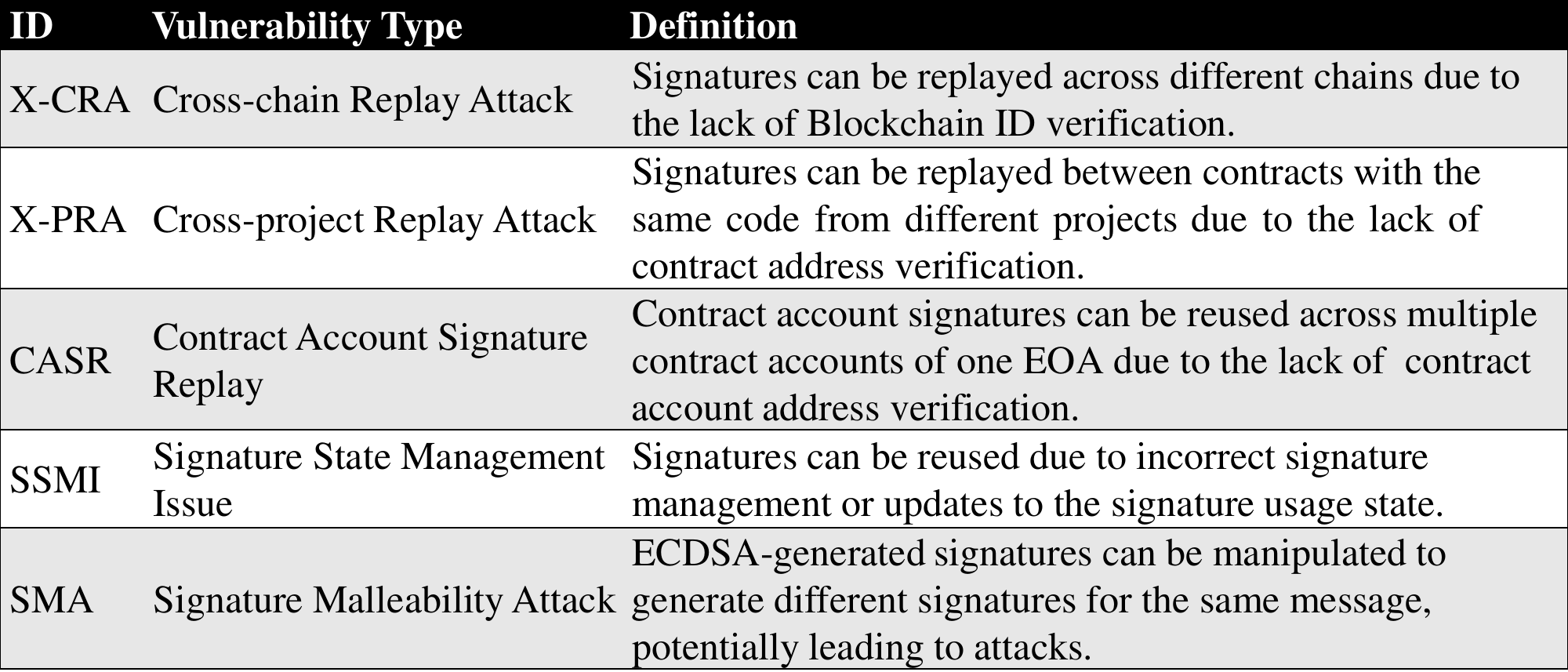}
    % \vspace{-1em}
    \caption{Definitions of Five Types of SRVs.}
   % \vspace{-0.5em}
    \label{fig:definetab}
\end{figure}

In this subsection, we define five types of SRVs, as shown in Figure~\ref{fig:definetab} with their corresponding \textit{Definitions} and \textit{IDs}. Each type is detailed and explained with \textit{Code Examples} from real-world audit reports.

% \vspace{-0.3cm}                           
\vspace{-0.3cm}

\paragraph{\textbf{(1) Cross-chain Replay Attack (X-CRA)}}
The \textit{Blockchain ID} is a unique identifier for distinguishing one blockchain from another~\cite{wood2014ethereum}. During signature verification, checking the \textit{Blockchain ID} in the signature message can restrict verification to a specific blockchain~\cite{eip4337}. The absence of \textit{Blockchain ID} checks in the signature message can lead to the same signature being reused across multiple blockchains, resulting in X-CRA.

\textbf{Code Example:} 
Figure~\ref{fig:ccra} illustrates the X-CRA identified by auditors in the \textit{Biconomy} project. The issue arises from the \textit{getHash()} function, not including \textit{block.chainid} in the signature message, invalidating blockchain-specific restrictions of the signature verification. Auditors noted that deploying this code on two blockchains would allow the same signature to be verified on both. They recommended adhering to the \textit{EIP-4337 standard}~\cite{eip4337}, which includes adding \textit{block.chainid} to the signature message to prevent X-CRA.

\vspace{-0.3cm}
\begin{figure}[h]
	\setlength{\abovecaptionskip}{-0.1cm}
	\begin{lstlisting}[xleftmargin=0.1cm,xrightmargin=0.2cm]
function getHash(UserOperation userOp) public {
     keccak256(abi.encode(
        block.chainid // @audit add chain id 
        ..., userOp.getSender(), userOp.nonce));
}
	\end{lstlisting}
	\caption{\textit{X-CRA} in \textit{Biconomy} Project~\cite{ccraexample}.}
	\label{fig:ccra}
\end{figure}
\vspace{-0.3cm}

\paragraph{\textbf{(2) Cross-project Replay Attack (X-PRA)}}
The contract address is a critical credential for distinguishing between projects deployed with the same code. Without contract address checks during the signature verification, the same signature can be reused by different projects. For instance, an attacker could reuse a valid signature from one project to execute unauthorized actions on another project.

\vspace{-0.3cm}
\begin{figure}[htb]
	\setlength{\abovecaptionskip}{-0.1cm}
	\begin{lstlisting}[xleftmargin=0.1cm,xrightmargin=0.2cm]
function _checkSig(...) public {
    bytes32 messageDigest = keccak256(..., 
      address(this));  // @audit add contract address
    ecrecover(messageDigest, v, r, s);
    ...
}
	\end{lstlisting}
	\caption{\textit{X-PRA} in \textit{Hermez Project}~\cite{cpraexample}.}
	\label{fig:cpra}
\end{figure}
\vspace{-0.3cm}

\textbf{Code Example:} Figure~\ref{fig:cpra} illustrates the X-PRA discovered by auditors in the \textit{Hermez} project. The \textit{\_checkSig()} function does not include the project address (\textit{address(this)}) when hashing the signature message. This omission allows the same signature to be verified across different \textit{Hermez} forked projects. For example, an attacker could reuse a signature verified by the \textit{Hermez} project in a forked instance (\textit{Hermez\_forked}) on the same blockchain to steal funds. To prevent this, it is recommended to include the project contract address in the signature message, ensuring that signatures cannot be reused in different projects.

\paragraph{\textbf{(3) Contract Account Signature Replay (CASR)}}
\textit{Contract Account Signature}, where the contract itself signs instead of an \textit{Externally Owned Account} (EOA) using its private key~\cite{contractsign,eoa}. This process follows the \textit{EIP-1271} standard~\cite{eip1271}, which allows an EOA to create multiple contract accounts and sign securely. However, if contract account address checks are omitted during signature verification, the same signature could be validated by different contract accounts, resulting in CASR.

\vspace{-0.3cm}
\begin{figure}[htb]
	\setlength{\abovecaptionskip}{-0.1cm}
	\begin{lstlisting}[xleftmargin=0.1cm,xrightmargin=0.2cm]
function execScheduled(Identity identity, bytes32 accHash, uint nonce, ... calldata txns) external {
     bytes32 hash = keccak256(..., address(identity),  // @audit add identity address 
          accHash, nonce, txns, false);
     require(scheduled[hash] != 0 && ...);
}
	\end{lstlisting}
	\caption{\textit{CASR} in \textit{AdEx protocol}~\cite{ade}.}
	\label{fig:casr}
\end{figure}
\vspace{-0.3cm}

\textbf{Code Example:} Figure~\ref{fig:casr} presents an example of CASR in the \textit{AdEx protocol}, which utilizes the \textit{EIP-1271} standard for \textit{QuickAccount}~\cite{quickacc} to manage multiple identities (contract accounts) and verify contract signatures. However, as shown in line 2 of Figure~\ref{fig:casr}, the signature message lacks the identity address, allowing it to be verified by multiple identities (contract accounts) and leading to asset loss. Auditors recommend incorporating the identity address into the signature message to prevent CASR.

\paragraph{\textbf{(4) Signature State Management Issue (SSMI)}}
To ensure effective signature state management, \textit{EIP-712}~\cite{eip712} introduces the \textit{domainSeparator}, which structurally organizes signature information such as \textit{timestamps} and \textit{nonces}, facilitating the verification and management of signature states. However, due to insufficient security awareness among developers, custom flawed signature management often results in disordered signature states. These vulnerabilities in signature state management can be exploited by attackers to perform replay attacks.

\vspace{-0.3cm}
\begin{figure}[htb]
	\setlength{\abovecaptionskip}{-0.1cm}
	\begin{lstlisting}[xleftmargin=0.1cm,xrightmargin=0.2cm]
function recoverSignature(...) returns (address) {
    // @audit Adherence to EIP-712
    SignedData memory payload = SignedData({
        transactionId: transactionId,... });
    return ECDSA.recover(ECDSA.toEthSignedMessageHash(keccak256(abi.encode(payload))), signature);
}
	\end{lstlisting}
	\caption{\textit{SSMI} in \textit{Connext NXTP}~\cite{connext}}
	\label{fig:ssmi}
\end{figure}
\vspace{-0.3cm}

\textbf{Code Example:} In Figure~\ref{fig:ssmi}, the \textit{recoverSignature()} function generates the custom signature message to verify the signer’s identity. However, the lack of a mechanism to track signature usage results in disordered signature states, leading to \textit{SSMI}. Auditors recommend strictly adhering to the \textit{EIP-712} during signature verification~\cite{connext}, including the use of a nonce to track and prevent signature reuse.

\paragraph{\textbf{(5) Signature Malleability Attack (SMA)}}
The \textit{Elliptic Curve Digital Signature Algorithm} (ECDSA) employed by \textit{Ethereum} is vulnerable to signature malleability attacks~\cite{signaturemall}. In such attacks, an attacker can modify specific parts of a signature without access to the private key, creating a new valid signature corresponding to the original signature message. To avoid such attacks, it is essential to enforce restrictions on the variables \textit{v} and \textit{s} when using the \textit{ecrecover(hash, v, r, s)} function. However, due to insufficient secure development expertise, developers often directly use \textit{ecrecover(hash, v, r, s)} for verification without implementing these checks.

\vspace{-0.3cm}
\begin{figure}[htb]
	\setlength{\abovecaptionskip}{-0.1cm}
	\begin{lstlisting}[xleftmargin=0.1cm,xrightmargin=0.2cm]
function permit(owner, spender, amount, v, r, s) {
    bytes32 permitDataDigest = keccak256(abi.encode(PERMIT_TYPEHASH, owner, spender));
    bytes32 digest = keccak256(abi.encodePacked("\x19\x01", DOMAIN_SEPARATOR(), permitDataDigest));
    require(owner == ecrecover(digest, v, r, s));
    allowances[owner][spender] = amount;
}
	\end{lstlisting}
	\caption{\textit{SMA} in \textit{Interest Protocol}~\cite{interest}.}
	\label{fig:sma}
\end{figure}
\vspace{-0.3cm}

\textbf{Code Example:} Figure~\ref{fig:sma} shows the SMA in the \textit{Interest Protocol}~\cite{interest}, where the \textit{permit()} function directly calls \textit{ecrecover()} for signature verification, exposing the protocol to replay attacks. Auditors recommend the following measures to mitigate this risk: 1) Ensure the \textit{s} value falls within the range \textit{$0 < s < secp256k1n \div 2 + 1$} (the lower half of the range). 2) Restrict the \textit{v} value to 27 or 28. Furthermore, adopting secure contract libraries, such as version 4.7.3 or later of \textit{OpenZeppelin}'s ECDSA library~\cite{openzeppelin}, ensures unique signature verification and prevents malleability attacks. These measures can effectively reduce the risk of signature malleability.

\section{Methodology}
In this section, we introduce \textit{LASiR}'s methodology, utilizing LLMs to understand contract semantics, combining static taint analysis and symbolic execution to enhance detection reliability.

\subsection{Overview}
\begin{figure}[htb]
    \centering
    % \vspace{-10pt}
    \includegraphics[width=\linewidth]{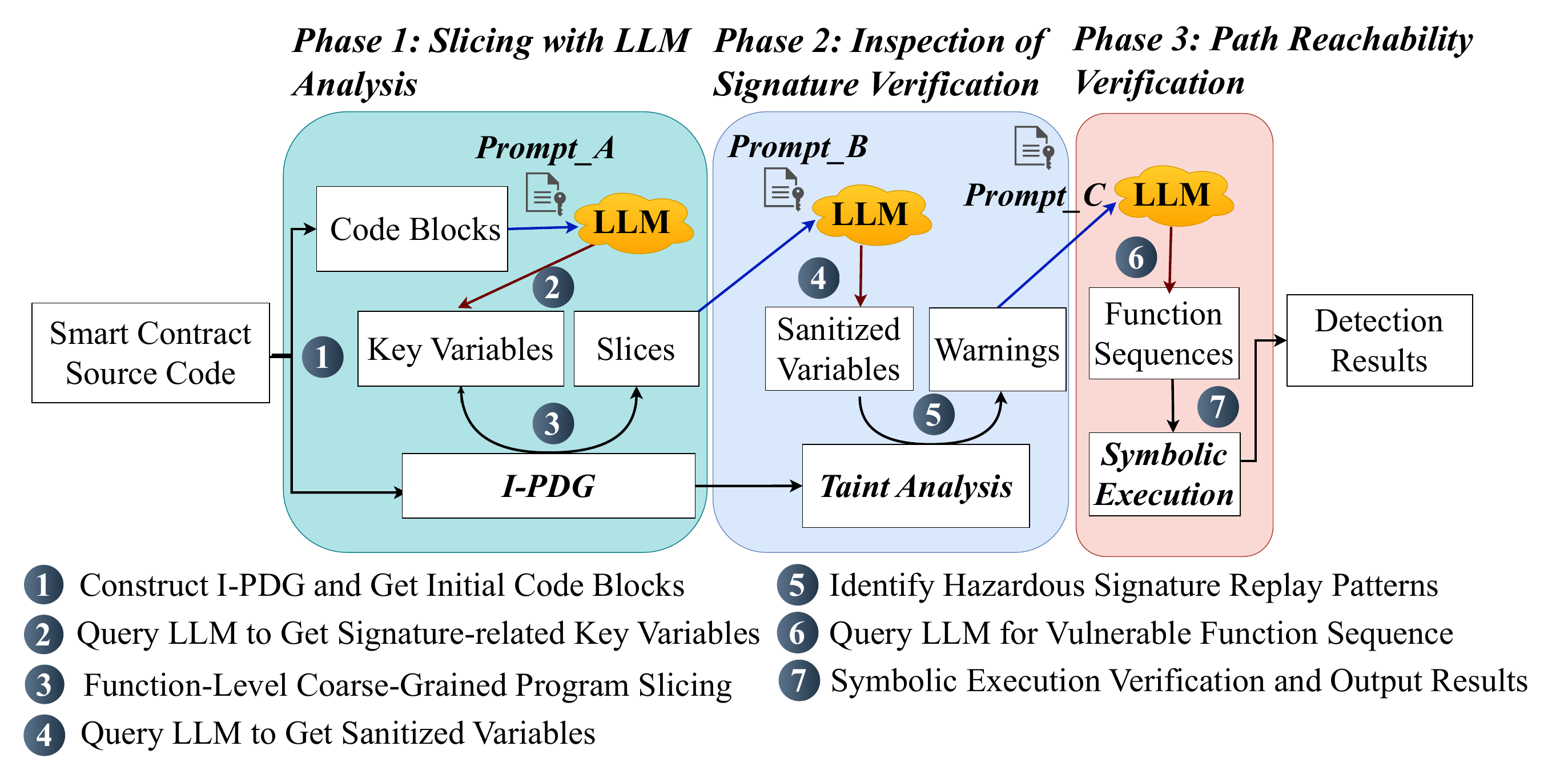}
    % \vspace{-1em}
    \caption{Overview of \textit{LASiR}.}
   % \vspace{-0.5em}
    \label{fig:Static4LLM}
\end{figure}

Figure~\ref{fig:Static4LLM} shows an overview of \textit{LASiR}, which inputs smart contract source codes and outputs the detection results. The detection process consists of three phases: \textit{Slicing with LLM Analysis}, \textit{Inspection of Signature Verification}, and \textit{Path Reachability Verification}.

\vspace{-6pt}
\vspace{-6pt}

\textit{\textbf{Phase 1: Slicing with LLM Analysis.}} To mitigate the impact of LLM input length limitations, we construct the \textit{Inter-contract Program Dependency Graph} (I-PDG)~\cite{wang2024efficiently} to comprehensively analyze dependencies and slice the code related to signature verification. The I-PDG represents global state dependencies, allowing for quick identification of dependencies between variables and statements. \textit{LASiR} integrates the \textit{Initial Code Block} that calls \textit{ecrecover()} with \textit{Prompt\_A} to query the LLM for key variables related to signature verification. Based on the dependencies of these key variables from the I-PDG, it slices the code that executes signature verification.

\textit{\textbf{Phase 2: Inspection of Signature Verification.}} \textit{LASiR} utilizes the general understanding capabilities of LLMs for automated state inspection, enhancing the precision of the static taint analysis. \textit{LASiR} uses \textit{Prompt\_B} to query the LLM for sanitized variables about the signature from sliced code. During taint analysis, it examines the dependencies of these variables to ensure sources are not contaminated by the time they reach sinks. By combining various \textit{domain-specific patterns} (details for Section~\ref{expertpatten}), it identifies risky behaviors and outputs relevant function names as \textit{Warnings}.

\textit{\textbf{Phase 3: Path Reachability Verification.}} To enhance detection reliability, \textit{LASiR} employs self-validation through symbolic execution to verify path reachability. \textit{LASiR} uses \textit{Prompt\_C} to instruct the LLM to understand the semantic context of functions from \textit{Warnings} and generate the sequence of functions containing risky logic or operations. Subsequently, symbolic execution is employed to explore execution paths derived from different function sequences, proving their reachability, and outputting detection results.

% In summary, \textit{LASiR} decouples the detection task into two aspects: semantic understanding and program verification. By leveraging the LLM's general semantic comprehension capabilities, it analyzes key variables in complex semantics while incorporating taint analysis and symbolic execution tools to enhance the reliability of signature replay vulnerability detection.

\subsection{Slicing with LLM Analysis}

\subsubsection{Step 1: Construct I-PDG and Get Initial Code Blocks.}
\begin{figure}[htb]
    \centering
    % \vspace{-10pt}
    \includegraphics[width=0.9\linewidth]{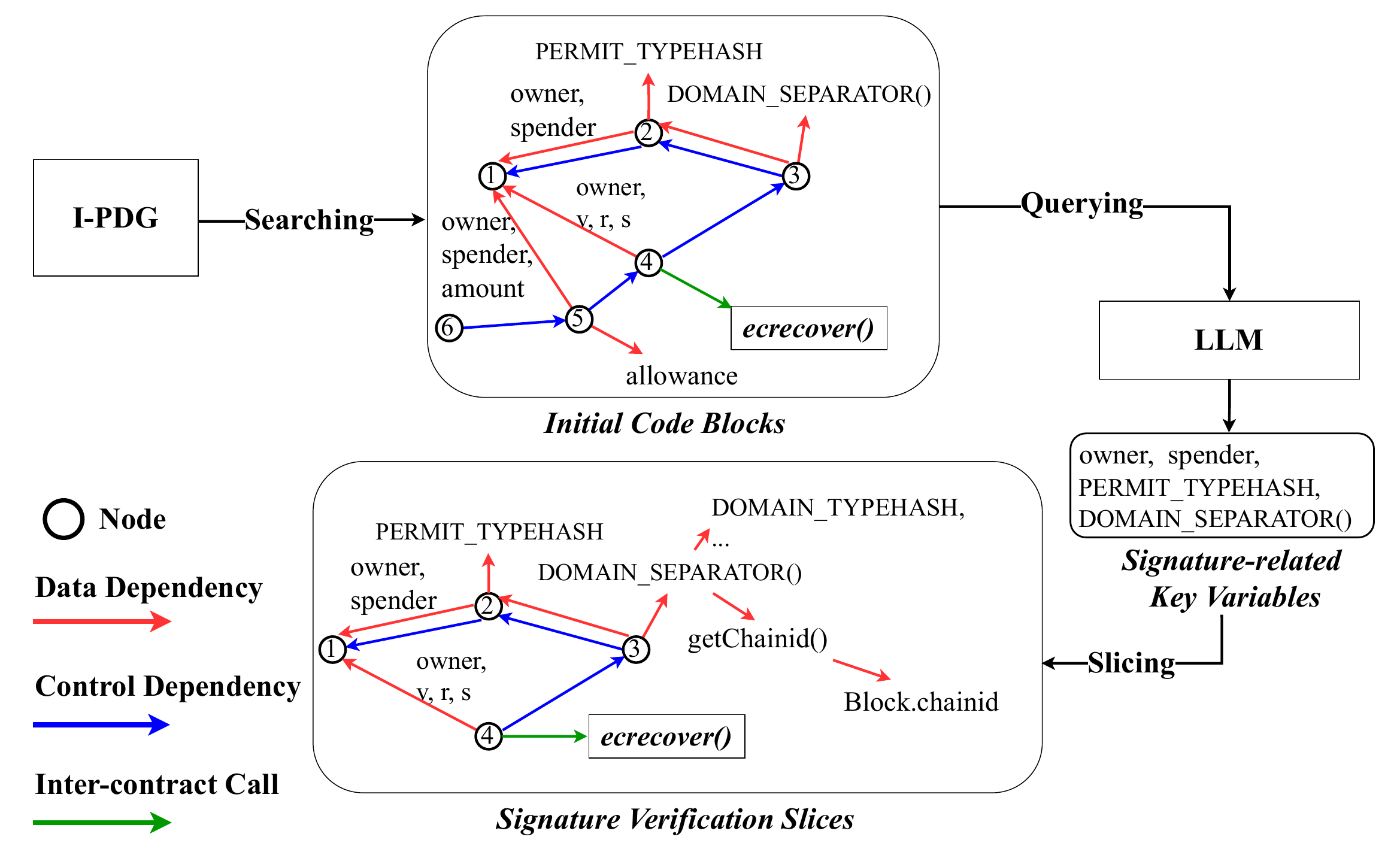}
    % \vspace{-1em}
    \caption{Slicing with LLM Analysis for \textit{Interest Protocol}~\cite{interest}.}
   % \vspace{-0.5em}
    \label{fig:slicing}
\end{figure}

\textit{LASiR} takes smart contract source code as input, compiles it into the \textit{Abstract Syntax Tree} (AST), and subsequently analyzes the AST to generate the \textit{Inter-contract Program Dependency Graph} (I-PDG)~\cite{wang2024efficiently}. The I-PDG integrates global control flow and inter-contract calls from the \textit{Inter-contract Control Flow Graph} (I-CFG)~\cite{pluto}, supplements data dependencies, and establishes global program dependencies. The nodes in the graph are derived from AST statements, while the edges represent data and control dependencies to model the program's structure. Specifically, as shown in Figure~\ref{fig:slicing}, the numbers in circles (nodes) correspond to line numbers in Figure~\ref{fig:sma}, with red, blue, and green lines representing data dependencies, control dependencies, and inter-contract calls, respectively. Figure~\ref{fig:slicing} illustrates the slicing analysis with LLMs for the \textit{Interest Protocol}. \textit{LASiR} collects all nodes within functions that contain the \texttt{ecrecover()} call as \textit{Initial Code Blocks} and further analyzes them using LLMs. We define Initial Code Blocks as the set of nodes that include the \texttt{ecrecover()} invocation along with all its dependent nodes. As shown in Figure~\ref{fig:slicing}, this includes not only lines 1–6 in Figure~\ref{fig:sma}, but also the \texttt{PERMIT\_TYPEHASH} state variable and the call to the \texttt{DOMAIN\_SEPARATOR()} function. All nodes related to the dependencies of \texttt{DOMAIN\_SEPARATOR()} are also included in the slice.

\vspace{-6pt}
\vspace{-6pt}

% \textit{LASiR} takes the smart contract source code as input, compiles it to generate the AST file, and constructs the \textit{Inter-contract Program Dependency Graph} (I-PDG) based on AST analysis. To analyze global program dependencies, we construct the I-PDG, which captures the global state dependencies of contracts. This allows for quick identification of dependencies between variables and statements. We utilize the \textit{Initial Code Block} containing the \textit{ecrecover()} function as the starting point for slicing, as it is essential for signature verification. Figure~\ref{fig:initalcode} presents an example of the \textit{Initial Code Block}, an independent function code block that does not include dependencies between different functions and contracts.

% \vspace{-0.4cm}
% \begin{figure}[htb]
% 	\setlength{\abovecaptionskip}{0.2cm}
% 	\begin{lstlisting}[xleftmargin=0.1cm,xrightmargin=0.2cm]
% function verifyEIP712(...) {
%     bytes32 hash = keccak256(abi.encodePacked("\x19\x01", DOMAIN_SEPARATOR, hashStruct));
%     address signer = ecrecover(hash, v, r, s);
%     // irrelevant variables
%     orderStatus[_guiltyOrderID] = OrderStatus.Slashed;
%     ...
% }
% 	\end{lstlisting}
% 	\caption{The example of \textit{Initial Code Block}}
% 	\label{fig:initalcode}
% \end{figure}
% \vspace{-0.4cm}

\subsubsection{Step 2: Query LLM to Get Signature-related Key Variables.} 
Static analysis often struggles to extract task-specific dependencies due to limited semantic understanding, impacting detection accuracy. To address this, we designed \textit{Prompt\_A} to guide the LLM in analyzing signature variables based on contract semantics and intent. These variables and their dependencies help static analysis focus on the signature status, thereby enhancing slicing accuracy.

We designed \textit{Prompt\_A} based on common practices~\cite{promptpractice} and \textit{Tier of Thought} (ToT) design~\cite{tot}. The analysis task is divided into three tiers for complex tasks, using outputs from previous tiers to generate responses for more challenging tasks, ensuring the reliability of the LLM's outputs. For structured output, the LLM returns results in \textit{JSON} format only in the final round to maintain its thought process. Specifically, \textit{Prompt\_A} guides the LLM to simulate a smart contract security auditor's workflow, aiding in accurately identifying and extracting key variables for signature verification through role-playing and structured analysis.

\begin{figure}[htb]
    \centering
    \vspace{-5pt}
    \includegraphics[width=0.98\linewidth]{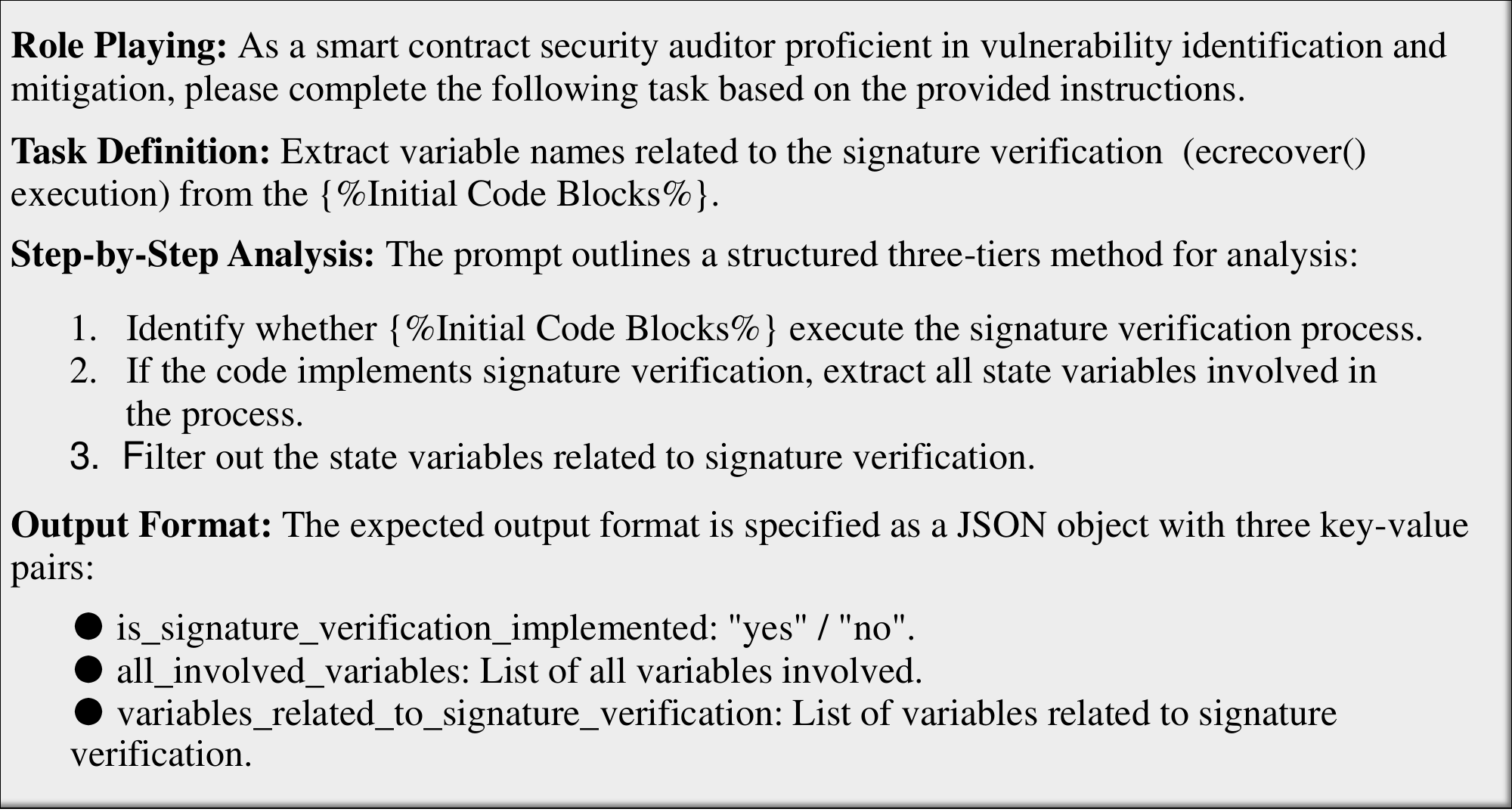}
    % \vspace{-1em}
    \caption{\textit{Prompt\_A} Template Design.}
   % \vspace{-1em}
    \label{fig:prompta}
\end{figure}

As shown in Figure~\ref{fig:prompta}, \textit{Prompt\_A} consists of four parts: \textit{Role Playing}, \textit{Task Definition}, \textit{Step-by-Step Analysis}, and \textit{Output Format}. In \textit{Role Playing}, we define the LLM as a smart contract security auditor skilled in identifying and mitigating vulnerabilities, activating its vulnerability analysis capabilities. The \textit{Task Definition} clearly states the task: extract variable names related to signature verification (\textit{ecrecover()}) from the \textit{{\%Initial Code Blocks\%}}. The \textit{Step-by-Step Analysis} breaks the task into three tiers: 1) Determine if the \textit{{\%Initial Code Blocks\%}} implements the signature verification. 2) If the code does implement signature verification, extract all state variables involved. 3) Filter out variables that affect signature verification. The \textit{Output Format} specifies that the output should be a \textit{JSON} object containing the required information. Since static analysis often retains irrelevant variables due to a lack of semantic context, it compromises accuracy. By leveraging the LLM, we can effectively obtain key variables related to signatures. As shown in Figure~\ref{fig:slicing}, when analyzing \textit{Signature-related Key Variables}, the LLM, with its semantic analysis, filters out irrelevant variables such as \textit{allowance} in line 5 of Figure~\ref{fig:sma} (whereas static analysis considers it due to control dependencies), thereby improving the accuracy.

\vspace{-0.1cm}
\vspace{-0.3cm}

\subsubsection{Step 3: Function-Level Coarse-Grained Program Slicing.}
To mitigate the limitations of LLM on input text length while maintaining state completeness, \textit{LASiR} employs function-level slicing based on program dependencies. It uses \textit{Signature-related Key Variables} from \textit{Step 2} and their dependencies to slice the code related to signature verification. Specifically, \textit{LASiR} treats each function as a complete unit rather than analyzing individual statements. If any statement within a function depends on \textit{Signature-related Key Variables}, the entire function is included. As shown in the \textit{Signature Verification Slices} in Figure~\ref{fig:slicing}, key variable \textit{DOMAIN\_SEPARATOR()} relies on \textit{Block.chainid} from the \textit{getChainid()} function, thus the \textit{getChainid()} function is included in the slice. \textit{LASiR} uses a coarse-grained, function-level program slicing approach~\cite{slicefunction} to avoid the fragmentation of functions and compromise of information integrity that often occurs with fine-grained slicing. This approach ensures that all relevant statements are preserved, enhancing the semantic understanding and contextual integrity of LLM analysis.

% To mitigate the limitations of long text inputs in LLMs while maintaining relevant state completeness, \textit{LASiR} employs coarse-grained slicing based on dependencies in contracts. It targets code related to signature verification and extracts comprehensive program slices. The key signature variables from \textit{Step 2}, along with their dependencies, aid in accurately slicing the signature verification code. 

% To prevent issues from fine-grained slicing, which can fragment functions and compromise information integrity for subsequent semantic analysis by LLMs, \textit{LASiR} adopts a function-level coarse-grained slicing approach~\cite{slicefunction}. This treats functions as the smallest slicing unit, resulting in a collection of functions related to signature verification. Slicing the program at the function level means treating each function as a complete unit instead of analyzing individual statements. This approach creates a set of functions impacting signature verification. If any statement within a function relies on key variables obtained in \textit{Step 2}, the entire function is included, ensuring all relevant statements are preserved and enhancing semantic understanding and analysis by LLMs.

\subsection{Inspection of Signature Verification}
% This subsection describes the process for checking the replay risk in signature verification.
\subsubsection{Step 4: Query LLM to Get Sanitized Variables.}
In static taint analysis, taint data is sanitized into variables without sensitive data. However, due to the complexity and variability of sanitization processes, existing static analyses heavily rely on fixed rules or manual inspection, limiting their effectiveness. To automate variable sanitization analysis, we leverage the LLM's understanding of contract semantics, transforming the task into natural language instructions through prompt design and identifying sanitized variables.

\begin{figure}[htb]
    \centering
    % \vspace{-5pt}
    \includegraphics[width=0.98\linewidth]{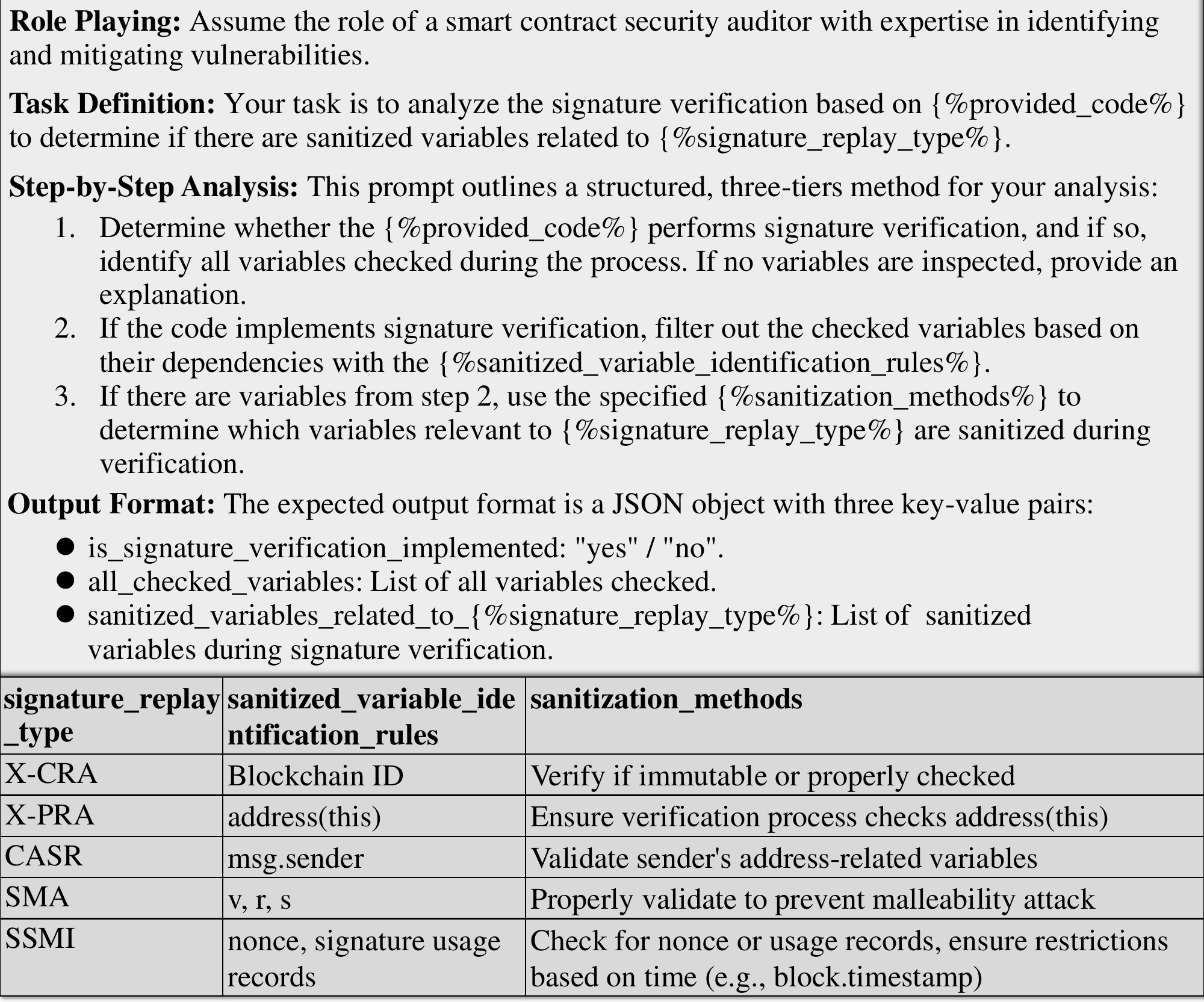}
    % \vspace{-1em}
    \caption{\textit{Prompt\_B} Template Design.}
   % \vspace{-1em}
    \label{fig:promptb}
\end{figure}

\sloppy
To guide the LLM in identifying sanitized variables effectively, we designed \textit{Prompt\_B}. As shown in Figure~\ref{fig:promptb}, the prompt outlines the LLM's task to analyze sanitized variables related to \textit{{\%sanitized\_variable\_type\%}} from the \textit{{\%provided\_code\%}}. The analysis is structured into three tiers: 1) search for all variables checked in signature verification and explain if none are found. 2) identify key variables based on their dependencies with the \textit{{\%sanitized\_variable\_identification\_rules\%}}. if signature verification is present. 3) If variables from tier 2 exist, use the specified \textit{{\%sanitization\_methods\%}} to determine which variables relevant to \textit{{\%signature\_replay\_type\%}} are sanitized during verification. The table below categorizes the identification rules and sanitization methods for five types of SRVs. \textit{{\%signature\_replay\_type\%}}, \textit{{\%sanitized\_variable\_identification\_rules\%}}, and \textit{{\%sanitization\_methods\%}} clarify the SRVs types, identification rules, and sanitization methods. This table guides the LLM in identifying sanitized variables and improving accuracy through targeted sanitization processes.

\vspace{-0.3cm}
\vspace{-0.5cm}

\subsubsection{Step 5: Identify Hazardous Signature Replay Patterns.}
The LLM-identified sanitized variable information from \textit{Step 4} lacks contract execution context and cannot be used directly for static taint analysis. To address this, \textit{LASiR} searches for related code blocks based on the program dependencies of the sanitized variables. By combining these related code blocks with \textit{domain-specific patterns}, it identifies risky signature patterns. Specifically, \textit{LASiR} iterates through all sanitized variables to find their definition nodes and performs a \textit{depth-first search} (DFS) of the entire I-PDG starting from these nodes. When dependencies on sanitized variables are found, the node is saved in \textit{Warning\_nodes}, and all successor nodes are recursively traversed. This generates a set of nodes that contain dependencies on sanitized variables (\textit{Warning\_nodes}), providing context for their operations. By combining these with \textit{domain-specific patterns} based on different SRVs analysis strategies (see subsection~\ref{expertpatten}), \textit{LASiR} identifies hazardous signature patterns and generates the corresponding function names as \textit{Warnings}.

\subsection{Path Reachability Verfication}
% In this subsection, we explain how \textit{LASiR} combines self-validation with symbolic execution to enhance the reliability of detection results.
\subsubsection{Step 6: Query LLM for Vulnerable Function Sequence.}
To ensure reliable results, \textit{LASiR} uses LLM's general understanding to review \textit{Warnings} from static taint analysis in \textit{Step 5}. \textit{LASiR} uses \textit{Prompt\_C} to guide the LLM in checking and correcting these \textit{Warnings}. If the \textit{Warnings} are confirmed, the LLM provides a function sequence related to the execution of the vulnerability. This sequence, which comprises multiple functions in a specific order, is generated by LLM reasoning. This information accelerates the traversal of the symbolic execution path, enhancing the efficiency of the analysis.

\vspace{-6pt}
\begin{figure}[htb]
    \centering
    % \vspace{-10pt}
    \includegraphics[width=0.98\linewidth]{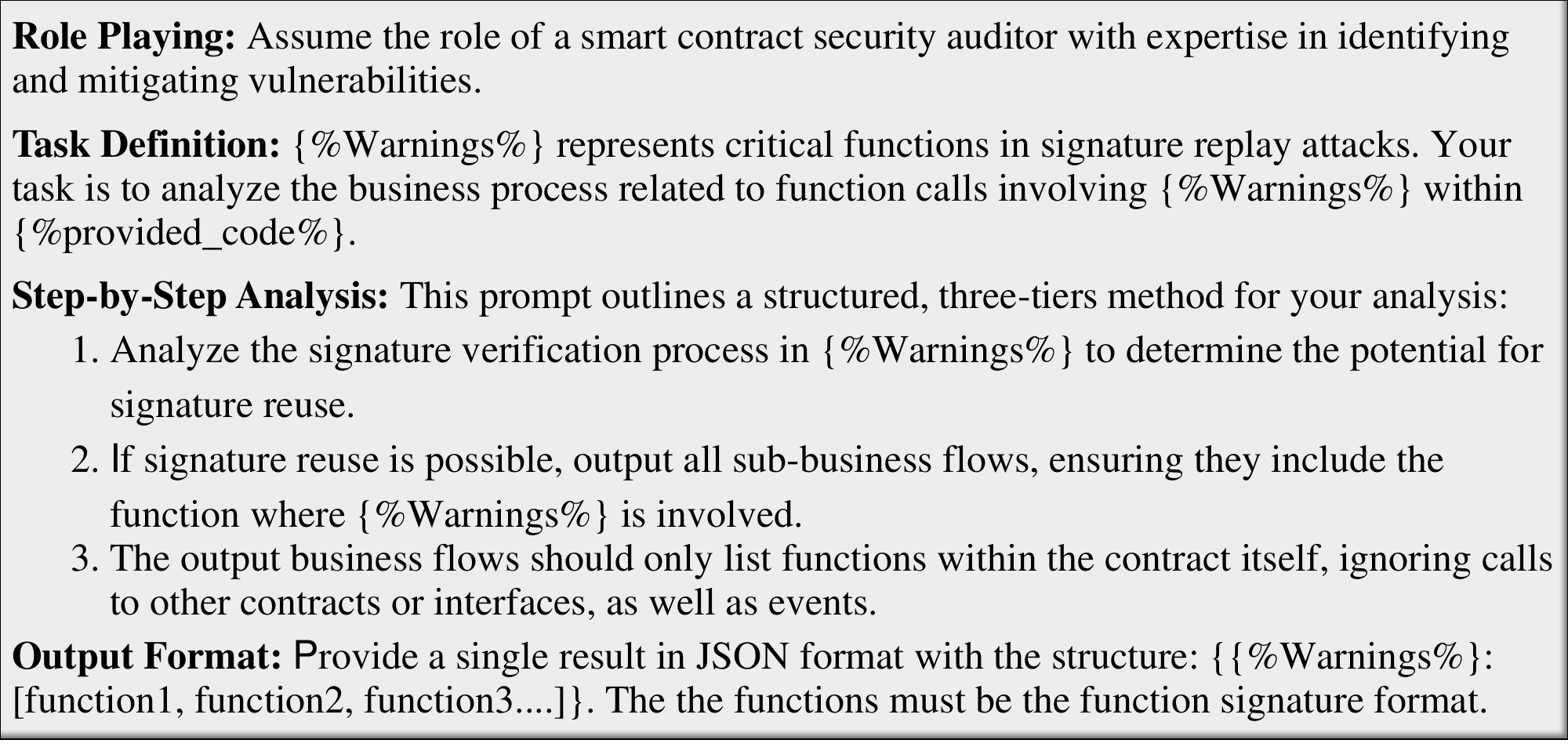}
    % \vspace{-1em}
    \caption{\textit{Prompt\_C} Template Design.}
   % \vspace{-1em}
    \label{fig:promptc}
\end{figure}
\vspace{-6pt}

Figure~\ref{fig:promptc} shows \textit{Prompt\_C} template. In the task definition, \textit{{\%Warnings\%}} is declared to be associated with signature replay attacks. The LLM's task is to analyze the business process related to function calls involving \textit{{\%Warnings\%}} within \textit{{\%provided\_code\%}}. A structured three-tier analysis method guides the LLM in reasoning about the vulnerability function sequence, 1) analyzing the signature verification process in \textit{{\%Warnings\%}} to determine the potential for signature reuse. 2) If signature reuse is possible, it should output all sub-business flows, ensuring \textit{{\%Warnings\%}} is involved. 3) Output business flows should only list functions within the contract itself, ignoring calls to other contracts or interfaces, as well as events. The output format specifies that the results should be in \textit{JSON} format with the structure: \textit{{{\%Warnings\%}: [function1, function2, function3,...]}}. This \textit{JSON} information contains the function sequence related to signature reuse. Utilizing the LLM's semantic understanding for pruning helps improve symbolic execution search efficiency.

% \vspace{-0.3cm}
% \begin{figure}[htb]
%     \centering
%     % \vspace{-10pt}
%     \includegraphics[width=0.9\linewidth]{./figures/symbolicexecution.pdf}
%     % \vspace{-1em}
%     \caption{Symbolic Execution for Path Reachability.}
%    % \vspace{-0.5em}
%     \label{fig:se4pr}
% \end{figure}
% \vspace{-0.3cm}

\subsubsection{Step 7: Symbolic Execution Verification and Output Results.}
To avoid static taint analysis causing permissions to be ignored or extracted incorrectly, and to mitigate misunderstandings by the LLM that affect reliability, \textit{LASiR} employs symbolic execution to verify path reachability. The function sequence from \textit{Step 6} guides the symbolic executor to traverse the CFG along different paths and collect permission-related path constraints. These constraints are then verified using SMT-based satisfiability checks to confirm path reachability. If the path constraints are solvable, indicating that the relevant permission checks can be passed, the vulnerability is proven to exist. This approach avoids blind searches and mitigates path explosions. Combined with the \textit{Warnings} identified by taint analysis, \textit{LASiR} verifies path reachability through symbolic execution and outputs the results.

\begin{figure}[h]
    \centering
    % \vspace{-10pt}
    \includegraphics[width=0.9\linewidth]{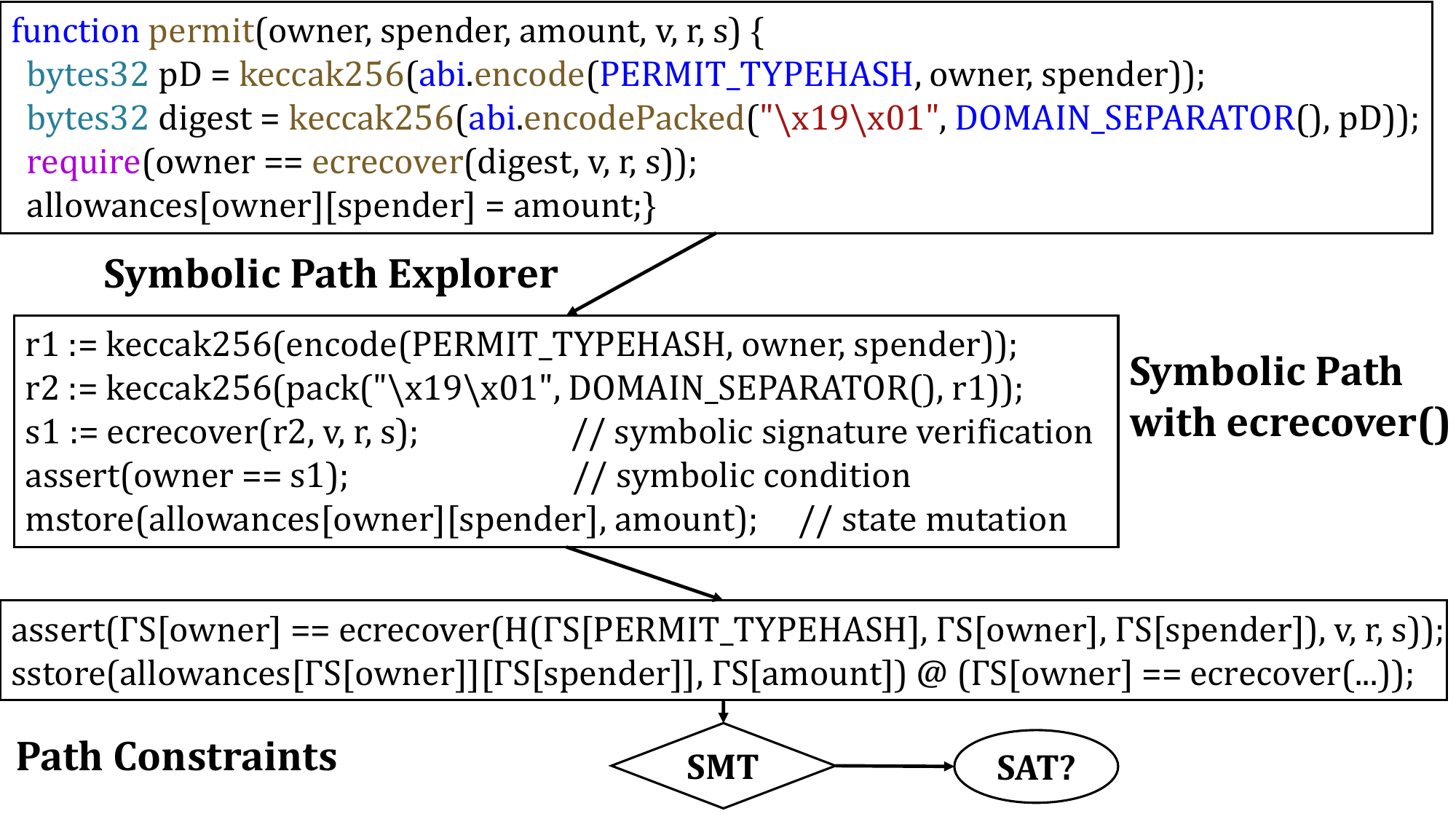}
    % \vspace{-1em}
    \caption{Symbolic Path Constraints for \texttt{permit()}.}
   % \vspace{-0.5em}
    \label{fig:se}
\end{figure}

\textit{LASiR} identifies feasible execution paths leading to \texttt{ecrecover()} based on CFG, and extracts both explicit and implicit control dependency conditions along these paths (e.g., \texttt{require(...)}, \texttt{if (msg.sender == owner)}). It then symbolically encodes the relevant variables by incorporating data dependencies, ultimately constructing complete path constraint expressions. Taking the \texttt{permit()} function illustrated in Figure~\ref{fig:se} as an example, the symbolic path includes signature digest construction (\texttt{digest = keccak256(...)}), signature recovery (\texttt{s = ecrecover(...)}), conditional checks (\texttt{require(owner == s)}), and state updates (\texttt{allowances[owner][spender] = amount}). The symbolic executor interprets the semantics of operations along the path and, using a symbolic memory model tailored to EVM instructions, generates corresponding path constraints. \textit{LASiR} then submits the combined path constraints to an SMT solver (e.g., Z3) to check for satisfiability. If the result is \texttt{SAT}, it indicates the existence of input parameters (e.g., \texttt{v}, \texttt{r}, \texttt{s}) that satisfy both the signature verification and access control conditions. In such cases, the path is deemed logically reachable, and a signature replay vulnerability is reported.

\vspace{-0.3cm}
\vspace{-0.3cm}

% To mitigate misunderstandings by the LLM and limitations in taint analysis (which affect reliability due to ignored or incorrectly extracted permissions), \textit{LASiR} employs symbolic execution to verify path reachability. Guided by the function sequence in \textit{Step 6}, it searches and traverses path constraints, using constraint solving to prove reachability. This approach effectively avoids blind searches and mitigates the path explosion. Combined with the signature replay vulnerability pattern identified by taint analysis, \textit{LASiR} proves their path reachability through symbolic execution and outputs the results.

% \vspace{-0.15cm}
\subsection{Signature Replay Attack Detection}
\label{expertpatten}
This subsection introduces \textit{domain-specific patterns} derived from various SRVs characteristics to identify hazardous patterns.

\textbf{\textit{Cross-chain Replay Attack (X-CRA).}} \textit{LASiR} analyzes the taint propagation path by examining variable reads and writes in \textit{Warnings\_nodes} to determine if any sanitization operations exist before reaching the sinks (\textit{ecrecover(\_hash, v, r, s)}). It specifically checks whether the \textit{\_hash} value is affected by \textit{block.chainid} and tracks the taint propagation of intermediate variables through program dependencies. It determines if the propagation is interrupted by any sanitization operations. If no sanitization is found and the \textit{\_hash} is contaminated, \textit{LASiR} identifies the X-CRA and generates \textit{Warnings}.

\textbf{\textit{Cross-Project Replay Attack (X-PRA).}} \textit{LASiR} analyzes variable dependencies in \textit{Warnings\_nodes} to confirm if the \textit{\_hash} value at the sinks (\textit{ecrecover(\_hash, v, r, s)}) is influenced by the \textit{address(this)} variable. Using the same detection logic as X-CRA, \textit{LASiR} employs program dependencies to check if the \textit{\_hash} value depends on \textit{address(this)} and verifies whether intermediate variables (influenced by \textit{address(this)}) have sanitization operations. If the taint propagation path is uninterrupted and there are no sanitization operations, \textit{LASiR} identifies the presence of X-PRA and generates \textit{Warnings}.

\textbf{\textit{Contract Account Signature Replay (CASR).}} \textit{LASiR} first retrieves the code blocks implementing the \textit{isValidSignature()} function~\cite{eip1271} from \textit{Warnings\_nodes}, as this function is crucial for contract signature verification. It then analyzes contamination at the \textit{ecrecover(\_hash, v, r, s)} sinks in these blocks, checking if the \textit{\_hash} value depends on \textit{msg.sender} information and whether intermediate variables (influenced by \textit{msg.sender}) are sanitized. If the taint propagation path is complete and lacks sanitization, it identifies the CASR risk and generates corresponding \textit{Warnings}.

\textbf{\textit{Signature State Management Issue (SSMI).}} To determine if the signature verification process includes verification of the signature usage state, \textit{LASiR} examines read and write operations affecting the \textit{\_hash} value before and after \textit{ecrecover(\_hash, v, r, s)}. Specifically, it checks for read-after-write operations for key variables identified in \textit{Step 4}. If these operations are not found, it concludes that there is no check for the signature usage state, identifying SSMI.

\textbf{\textit{Signature Malleability Attack (SMA).}} 
\textit{LASiR} analyzes the code blocks preceding \textit{ecrecover(\_hash, v, r, s)} to check for statements restricting the \textit{v} and \textit{s }variables. Ensures that \textit{$0 < s < secp256k1n \div 2 + 1$} and \textit{v} = 27 or 28. If these conditions are not met, SMA exists.
% \textit{LASiR} checks code before \textit{ecrecover(\_hash, v, r, s)} to ensure \textit{$0 < s < secp256k1n \div 2 + 1$} and \textit{v} = 27 or 28. If these conditions are not met, SMA exists.

\vspace{-0.15cm}
\section{Evaluation}
In this section, we analyze and evaluate the effectiveness of \textit{LASiR} in detecting SRVs by answering the following research questions:

\begin{itemize}
\setlength{\itemindent}{-6mm}
\item RQ1: How does \textit{LASiR} perform on the large-scale dataset?
\item RQ2: What is the performance of \textit{LASiR} in detecting SRVs?
\item RQ3: How does LLM enhance \textit{LASiR}'s effectiveness?
\end{itemize}

\subsection{Experiment Setup}
The experiment was carried out on a Ubuntu 20.04.1 LTS server equipped with a 16-core Intel(R) Xeon(R) Gold 5217 processor.

\textbf{Implementation:} \textit{LASiR} is implemented in Python, utilizing \textit{SlithIR}~\cite{slithir} for constructing the I-PDG and dependency analysis, and \textit{Rattle}~\cite{rattle} for Control Flow Graph (CFG) path recovery and symbol execution. Additionally, \textit{LASiR} is developed and tested using the \textit{DeepSeek-V3}~\cite{deepseekv3} LLM API services, known for its robust capabilities in code semantic tasks, offering a 128K input limit and low token price~\cite{deepseek_v3_release}. It employs default parameters, sets \textit{Temperature} to 0 to reduce randomness, and uses a three-request mechanism to ensure result stability by handling abnormal returns.

\textbf{Datasets:} To evaluate \textit{LASiR}'s performance, we selected \textit{Ethereum}~\cite{ethereum}, \textit{BSC}~\cite{bsc}, \textit{Polygon}~\cite{polygon}, and \textit{Arbitrum}~\cite{arbitrum} based on their \textit{Total Value Locked} (TVL) and active user rankings from \textit{DefiLlama}~\cite{defillama}, and collected 918,964 contract source codes (2017.10–2024.01). We then constructed two datasets to assess \textit{LASiR}'s effectiveness in detecting SRVs. \textbf{DB1:} To evaluate \textit{LASiR}'s performance on large-scale, real-world datasets, we searched for contracts containing \textit{ecrecover()} in their ASTs to identify those involving signature verification. This process yielded 15,383 contracts, distributed across \textit{Ethereum} (4,513), \textit{BSC} (5,590), \textit{Polygon} (4,140), and \textit{Arbitrum} (1,140). \textbf{DB2:} To analyze \textit{LASiR}'s accuracy, we manually analyzed 500 randomly selected contract source codes from \textit{DB1}, identifying 72 positive and 428 negative cases.

% \begin{table}[htb]
% \centering
% \caption{Sources and Composition of Contracts in DB1.}
% \vspace{-5pt}
% \begin{tabular}{l|llll}
% \hline
%  & \textbf{Ethereum} & \textbf{BSC} & \textbf{Polygon} & \textbf{Arbitrum} \\
% \hline
% \textbf{\# Contracts} & 339,341 & 366,086 & 150,816 & 62,721 \\
% \textbf{\# Contracts with Signatures} & 4,514 & 4,337 & 5,304 & 1,228 \\
% \hline
% \end{tabular}
% \label{tab:bg1}
% \end{table}

\vspace{-0.15cm}
\subsection{RQ1: Performance on large-scale dataset}
To evaluate \textit{LASiR}'s performance on large-scale datasets, we conducted experiments on \textit{DB1} dataset. During the experiment, we recorded statistics on the number, proportion, average detection time, and cost across four blockchains. 

\vspace{-6pt}
\begin{table}[H]
    \centering
    \large
    \caption{Statistics of Large-Scale Detection Results.}
    \vspace{-5pt}
    \resizebox{\linewidth}{!}{%
    \fontsize{20}{24}\selectfont
\begin{tabular}{l|llllllllll}
\hline
   & \textbf{X-CRA} & \textbf{X-PRA} & \textbf{CASR} & \textbf{SSMI} & \textbf{SMA} & \textbf{Proportion} & \textbf{Avg. Time (s)} & \textbf{Total Cost (\$)} \\ \hline
\textbf{Ethereum} & 455   & 493   & 14   & 734  & 674         & 19.63\%    & 38.27            & 4.40      \\
\textbf{BSC}      & 211   & 198   & 3    & 366  & 252         & 9.29\%     & 40.67            & 4.23      \\
\textbf{Polygon}  & 324   & 301   & 3    & 375  & 353         & 7.11\%     & 41.37            & 5.17      \\
\textbf{Arbitrum} & 51    & 52    & 2    & 73   & 68          & 5.94\%     & 40.94            & 1.20      \\ \hline
\end{tabular}
    }
    \label{tab:rq1}
\end{table}
\vspace{-6pt}

The detailed results in Table~\ref{tab:rq1} show that the higher quantities of SSMI and SMA pose significant threats to signature security. While CASR is less common, largely due to the adoption of the \textit{EIP-1271} standard~\cite{eip1271}, which enhances the security of contract account signatures. Notably, 19.63\% of contracts that use signatures on \textit{Ethereum} contain SRVs, highlighting a widespread issue. \textit{LASiR}'s average detection time is approximately 40 seconds, with an overall LLM API cost of \$15, showcasing its efficiency and cost-effectiveness for large-scale dataset detection.

% To analyze the real-world impact of SRVs, we analyzed the assets of contract addresses and manually assessed the risk of signature reuse. For effective analysis and balance of manual inspection efforts, we filtered contracts with balances greater than \$200, totaling 258 addresses. Further manual analysis found that 31 contracts' signature reuse can be exploited through forked blockchains, totaling \$4.76 million in active assets. 19 of these contracts could reuse signatures by modifying transaction input parameters (specific online signature replay links in our \href{https://anonymous.4open.science/r/LASiR-B207/Experimental_data/RQ1/Manual%20analysis.csv}{repository}).

To assess the real-world impact of signature reuse vulnerabilities (SRVs), we conducted an analysis of contract addresses with non-trivial asset holdings and manually evaluated their susceptibility to signature reuse. To ensure both meaningful coverage and practical feasibility, we filtered for contracts with non-zero balances, yielding a dataset of 258 contract addresses. Subsequent manual inspection revealed that 31 of these contracts exhibited signature reuse behaviors that could be exploited on forked blockchains, collectively securing approximately \$4.76 million in active assets. Among these, 24 contracts contained signature-verifiable transactions that were replayable by modifying input parameters. We validated their replayability using Tenderly’s online simulator, which allowed us to emulate transaction behavior. The remaining seven contracts required the manual construction of proof-of-concept (PoC) exploits to confirm their exploitability. All corresponding simulation links and PoC scripts are publicly available in our code repository. For ethical considerations, our validation strictly focused on confirming the presence of signature reuse behaviors without performing any asset-draining operations.

\textbf{Answer to RQ1:} \textit{LASiR} demonstrates rapid detection and cost efficiency in large-scale analyses. Experimental results show that 19.63\% of contracts using signatures contain SRVs on \textit{Ethereum}, with a wide distribution. Manual inspection further reveals that these vulnerabilities affect active assets worth \$4.76 million. 

\subsection{RQ2: Detection Performance Analysis}
To evaluate the performance of \textit{LASiR} in detecting SRVs, we conducted experiments with \textit{DB2} (including 72 positive and 428 negative labels). Furthermore, to analyze the effectiveness of \textit{LASiR}'s LLM-assisted static taint analysis, we conducted comparative experiments with existing tools, covering static analysis, LLM analysis, and the combined approach of LLM and static analysis.  

As shown in Table~\ref{tab:rq3}, \textit{LASiR} achieved a \textit{Precision} of 82.14\%, \textit{Recall} of 95.83\%, and an F1-score of 88.46\%, effectively detecting SRVs. Meanwhile, we conduct a further analysis of the reasons behind the 15 false positives (FP) and 3 false negatives (FN).

\textbf{False positives:} The primary causes originate from two aspects: Unrelated States Restrictions and Implementation Errors. Our analysis reveals that 11 FPs are due to state restrictions that are unrelated to signature verification. As shown in Figure~\ref{fig:fp}, while the \textit{get()} function includes signature checks, resulting in \textit{SMA}, the balance reset logic (\textit{$wallets[signer] != 0$}) at line 4 prevents token transfers, even if signature replay occurs. This also relates to \textit{LASiR}'s strategy of exclusively extracting signature-related states through static taint analysis to filter excessive divergence in LLM output, thereby overlooking unrelated signature state constraints that indirectly limit replay occurrence. Furthermore, the remaining 4 FPs involve developer-customized errors, including flawed verification execution and improper security library usage. Despite meeting SRVs definitions, signatures cannot be verified due to implementation errors, and such developer-induced errors are frequent.

\vspace{-0.3cm}
\begin{figure}[htb]
	\setlength{\abovecaptionskip}{0.2cm}
	\begin{lstlisting}[xleftmargin=0.1cm,xrightmargin=0.2cm]
function get(bytes32 _r, bytes32 _s, uint8 _v) {
    require(_v == 27 || _v == 28);
    address signer = ecrecover(..., _v, _r, _s);
    require(... && wallets[signer] != 0);
    payable(msg.sender).transfer(wallets[signer]);
    wallets[signer] = 0;
}
	\end{lstlisting}
	\caption{Unrelated States Restrictions.}
	\label{fig:fp}
\end{figure}
\vspace{-0.3cm}

\textbf{False Negatives:} Semantic analysis of the assembly code is limited, affecting the accuracy of the taint propagation. In Figure~\ref{fig:fp1}, the \textit{transfer\_from()} function parses signature data using the assembly code and verifies it through \textit{ecrecover()}. \textit{LASiR} identifies \textit{ecrecover()} as a sink in static taint analysis, tracking taints from the source (\textit{\_data}) to the sink. However, due to the LLM's lack of in-depth learning of smart contract assembly code, it often misinterprets the semantic functions of assembly code, resulting in incorrect taint tracking and subsequently causing false negatives.

\vspace{-0.3cm}
\begin{figure}[htb]
	\setlength{\abovecaptionskip}{0.2cm}
	\begin{lstlisting}[xleftmargin=0.1cm,xrightmargin=0.2cm]
function transfer_from(bytes memory _data) public  {
    assembly { sig_r := mload(_data); 
        sig_s := mload(add(_data, 32)); 
        sig_v := ... }
    ecrecover(limit_hash, sig_v, sig_r, sig_s );
    
	\end{lstlisting}
	\caption{Custom Assembly Implementation of \textit{ecrecover()}.}
	\label{fig:fp1}
\end{figure}
\vspace{-0.3cm}

\textbf{Comparison with existing tools.}
To further evaluate the effectiveness of the LLM-assisted static taint analysis of \textit{LASiR}, we extended existing general-purpose detection tools (covering static analysis, LLM analysis, and the combination of LLM analysis and static verification) with the definition of SRVs for comparison.

\vspace{-6pt}
\begin{table}[htb]
    \centering
    \large
    \caption{Comparison with Existing Tools}
    \vspace{-5pt}
    \resizebox{\linewidth}{!}{%
    \fontsize{20}{24}\selectfont
\begin{tabular}{l|lll|ll|ll}
\hline
\textbf{}                & \textbf{GPT-4o} & \textbf{DeepSeek-R1} & \textbf{DeepSeek-V3} & \textbf{Slither4SRV} & \textbf{Siguard} & \textbf{GPTScan} & \textbf{LASiR} \\ \hline
\textbf{TP}              & 36              & 50                   & 1                    & 61                   & 3                & 23               & 69             \\
\textbf{FP}              & 222             & 245                  & 7                    & 375                  & 0                & 162              & 15             \\
\textbf{FN}              & 36              & 22                   & 73                   & 11                   & 69               & 49               & 3              \\
\textbf{TN}              & 206             & 183                  & 419                  & 53                   & 418              & 266              & 413            \\ \hline
\textbf{Precision}       & 13.95\%         & 16.95\%              & 12.50\%              & 13.99\%              & 100.00\%         & 12.43\%          & 82.14\%        \\ \hline
\textbf{Recall}          & 50.00\%         & 69.44\%              & 1.35\%               & 84.72\%              & 4.17\%           & 31.94\%          & 95.83\%        \\ \hline
\textbf{F1-score}        & 21.82\%         & 27.25\%              & 2.44\%               & 24.02\%              & 8.00\%           & 17.90\%          & 88.46\%        \\ \hline
\textbf{Total Cost (\$)} & 17.01           & 4.25                 & 2.82                 & 0                    & 0                & 0.16             & 0.3            \\ \hline
\end{tabular}
    }
    \label{tab:rq3}
\end{table}
\vspace{-6pt}

To select general-purpose tools, we reviewed the top journals, conference papers, and audit reports to serach general analysis tools (e.g., ~\cite{slithir,securify,eTainter}). For static analysis tools, we selected \textit{Slither} for its scalable architecture and modular detection framework, thereby extending the development of \textit{Slither4SRV} with SRVs detection rules. Additionally, we incorporated Zhang et al.'s tool, \textit{Siguard}, which supports the detection of SSMI vulnerabilities and represents the first work targeting signature-related vulnerabilities~\cite{siguard}.  For LLM analysis, the SRVs definitions were structured as detection prompts and directly analyzed using three LLMs: \textit{GPT-4o}~\cite{chatgpt}, \textit{DeepSeek-R1}~\cite{deepseekr1}, and \textit{DeepSeek-V3}~\cite{deepseekv3}. \textit{GPT-4o} reflected benchmark levels, \textit{DeepSeek-R1} excelled in reasoning for complex tasks, and \textit{DeepSeek-V3} served as the base model for \textit{LASiR}, facilitating a comparison of \textit{LASiR} with pure LLM analysis. Additionally, \textit{GPTScan}~\cite{gptscan} was chosen for its combined approach of LLM analysis and static verification, with reconstructed properties to support SRVs analysis.

As shown in Table~\ref{tab:rq3}, \textit{LASiR} demonstrated outstanding performance, particularly in \textit{Precision} (82.14\%) and \textit{Recall} (95.83\%). In contrast, other tools revealed insufficient and unbalanced capabilities. For instance, \textit{Siguard} attained an \textit{F1-score} of only 8\%, detecting three true positives vulnerabilities but missing most SRVs, resulting in a \textit{Recall} of just 4.17\%. While \textit{Slither4SRV} achieved a high \textit{Recall} of 84.72\%, its \textit{Precision} was only 13.99\%, reflecting the limitations of pattern-based static analysis. Among the three LLM models, \textit{DeepSeek-R1} performed best with a \textit{Recall} of 69.44\% owing to its semantic understanding and reasoning abilities, but it also had the highest number of false positives (245), leading to a \textit{Precision} of 16.95\%. This poor performance was mainly due to LLM's limited program analysis capabilities and unstable output. Additionally, \textit{GPTScan4SRV} reduced false positives by verifying the LLM output by static analysis verification. However, the separation of LLM analysis from static analysis introduces challenges. Specifically, the reliance on fixed patterns in static analysis, coupled with inconsistencies in the LLM's interpretation of scenarios and properties, frequently leads to semantic information loss, which adversely affects detection performance. Comparison with existing tools shows \textit{LASiR} efficiently leverages the semantic understanding of the LLM to guide static taint analysis, using code syntax rules to filter LLM outputs in real time. For instance, while \textit{DeepSeek-V3} had a \textit{Recall} of 1.35\% and an F1-score of 2.44\%, \textit{LASiR}'s F1-score reached 88.46\% when using \textit{DeepSeek-V3} for static taint analysis, a 36-fold improvement. This approach maximizes the strengths of different technologies, providing a robust solution for SRVs detection.

\textbf{Aanswer to RQ2:} \textit{LASiR} detects SRVs leveraging the semantic understanding of LLM to aid in static taint analysis. Compared to existing methods, its \textit{F1-score} for SRVs detection reached 88.46\%, demonstrating excellent performance.

\subsection{RQ3: Impact of LLM on Performance}
To analyze the impact of LLM on enhancing \textit{LASiR}'s detection performance, we conducted ablation experiments with four groups based on the participation of LLM in different phases of detection. The groups were divided as follows: \textit{LLM\_Phases1\&2\&3} (participated in all three phases), \textit{LLM\_Phases1\&2} (participated in \textit{Phases 1 and 2} only), \textit{LLM\_Phase1} (participated in \textit{Phase 1} only), and \textit{No\_LLM} (did not participate in any phase). The experimental data, \textit{DB2}, included 500 contracts with 72 positive and 428 negative labels.

\vspace{-6pt}
\begin{table}[h]
    \centering
    \large
    \caption{Statistics of LLM's Impact on Different Phases}
    \vspace{-5pt}
    \resizebox{\linewidth}{!}{%
    \fontsize{20}{24}\selectfont
\begin{tabular}{l|llllllll}
\hline
                 & \textbf{TP} & \textbf{FP} & \textbf{TN} & \textbf{FN} & \textbf{Precision} & \textbf{Recall} & \textbf{F1-score} & \textbf{Time (s)} \\ \hline
\textbf{LLM\_Phases1\&2\&3} & 69          & 15          & 413         & 3           & 82.14\%            & 95.83\%         & 88.46\%     & 41.03    \\
\textbf{LLM\_Phases1\&2} & 35          & 45          & 383         & 37 $\uparrow$          & 43.75\% $\downarrow$           & 48.61\% $\downarrow$        & 46.05\% $\downarrow$     & 71.35 $\uparrow$   \\
\textbf{LLM\_Phase1} & 31          & 351 $\uparrow$        & 77          & 41          & 8.12\% $\downarrow$            & 43.06\%         & 13.66\%     & 55.27    \\
\textbf{No\_LLM} & 19 $\downarrow$         & 413         & 15          & 53          & 4.40\% $\downarrow$            & 26.39\% $\downarrow$        & 7.54\% $\downarrow$     & 15.49 $\downarrow$   \\ \hline
\end{tabular}
    }
    \label{tab:rq2}
\end{table}
\vspace{-6pt}

Table~\ref{tab:rq2} shows the statistical results of LLM participation in different phases of \textit{LASiR}, highlighting significant effects in each phase. Comparing \textit{LLM\_Phases1\&2\&3} with \textit{LLM\_Phases1\&2}, we observe that without LLM guidance in \textit{Phase 3}, detection metrics decrease by about half. This is because, during symbolic execution, the lack of LLM guidance greatly expands the path search space, causing many irrelevant functions to be traversed, and leading to many missed vulnerabilities and a significant increase in FNs. Comparing \textit{LLM\_Phases1\&2} with \textit{LLM\_Phase1}, the absence of LLM in \textit{Phase} 2's sanitization checks results in an incomplete state at the sinks, increasing FPs and dropping \textit{Precision} from 43.75\% to 8.12\%. These results underscore LLM's significant role in different phases.

Furthermore, \textit{LASiR} effectively leverages the semantic understanding of LLM to improve detection accuracy and completeness. Comparing \textit{No\_LLM} with \textit{LLM\_Phases1\&2\&3}, \textit{Precision} improved from 4.40\% to 82.14\%, \textit{Recall} from 26.39\% to 95.83\%, and \textit{F1-score} from 7.54\% to 88.46\%. These results showcase LLM's effectiveness in improving vulnerability detection accuracy, reducing false positive rates, and narrowing the symbolic execution search space.

To evaluate the LLM's impact on detection efficiency, we compared detection times in four groups. The results of Table~\ref{tab:rq2} showed that with LLM throughout the process (\textit{LLM\_Phases1\&2\&3}), detection time was 41.03 seconds, demonstrating balanced performance. Without LLM guidance from \textit{Phase 3} (\textit{LLM\_Phases1\&2}), detection time increased to 71.35 seconds due to symbolic execution path explosion. Without LLM support (\textit{No\_LLM}), most dangerous paths were ignored, reducing the detection time to 15.49 seconds but significantly lowering the accuracy and reliability.

\textbf{Aanswer to RQ3:} The semantic information from LLM, through contract context analysis, is crucial for \textit{LASiR}'s detection process. This enhances static taint analysis by contract semantic understanding, improving accuracy and efficiency.

\vspace{-0.15cm}
\section{Discussion}
\subsection{Case Study}

% \vspace{-0.15cm}
% \begin{figure}[htb]
%     \centering
%     % \vspace{-10pt}
%     \includegraphics[width=0.88\linewidth]{./figures/whitelist_attack.pdf}
%     % \vspace{-1em}
%     \caption{Misuse of Whitelist Signatures in \textit{Sol Flowers x AP}.}
%    \vspace{-1em}
%     \label{fig:casestudy}
% \end{figure}
% \vspace{-0.15cm}

We classify this case as \textit{Whitelist Signature Permission Abuse} because these NFT contracts incorrectly verify whitelist signatures, allowing a single signature to mint multiple NFTs, with assets totaling \$24,838. The \textit{Sol Flowers x AP} NFT contract~\cite{solflowers} allows whitelisted users to mint multiple NFTs using the same signature. The process is as follows:
\ding{182} The attacker completes an offline task. \ding{183} After completing the task, the attacker requests a signature to mint an NFT. \ding{184} The attacker obtains the signature. \ding{185} The attacker submits a minting request to the contract and mints an NFT. \ding{186} The attacker reuses the signature to submit another minting request and mints another NFT. Due to the contract's lack of signature usage management (i.e., it does not check if the signature has been verified), there is SSMI. This allows the attacker to reuse signatures to mint multiple NFTs, including high-rarity ones. This signature replay vulnerability severely impacts the assets of the \textit{Sol Flowers x AP}.

% We identified a significant number of NFT sales contracts with active asset holdings that improperly implement signature verification mechanisms, thereby exposing them to replay attacks. Our analysis reveals a systemic vulnerability originating from the widespread reuse of insecure contract templates, which propagates signature-related flaws across multiple NFT projects, particularly within the rapidly evolving and loosely standardised NFT ecosystem. To support reproducibility and further research, we have made the relevant exploit scripts publicly available in our code repository~\cite{gitrepository}. We explicitly declare that no asset draining actions were performed during the course of this study. A detailed examination of these exploitation cases is presented in Part~\uppercase\expandafter{\romannumeral3} of \textit{Appendix~A}.

\vspace{-0.15cm}
\subsection{Threats to Validity}
\textbf{Internal Validity:} 
% One internal threat lies in the information asymmetry between static analysis and LLM. To ensure a structured output and further analysis, key variables in the LLM output may lack the context required for static analysis. To ensure the integrity of static analysis, \textit{LASiR} analyzes the read/write operations and dependencies of key variables in the I-PDG, collecting the code blocks involved to provide the necessary context. Another internal threat is directing the LLM to focus only on states related to signature verification for consistent output, which can lead to false positives due to missing \textit{unrelated signature status restrictions}. To mitigate this, \textit{LASiR}’s self-verification in \textit{Phase 3} requires the LLM to review previous responses and further use symbolic execution to verify the reachability of the path, ensuring consistency and reliability.
One internal threat lies in the information asymmetry between static analysis and LLM. To ensure a structured output and further analysis, key variables from the LLM output may lack the context required for static analysis. To mitigate this, \textit{LASiR} analyzes dependencies of key variables within the I-PDG, gathering the relevant code blocks to provide necessary context. Furthermore, in \textit{Phase 3}, \textit{LASiR} conducts self-verification using symbolic execution to confirm path reachability, ensuring reliability. Another internal threat is the extension of \textit{Slither} and \textit{GPTScan} to support SRVs detection. After reviewing the state-of-the-art tools, we selected them because they are general-purpose tools with excellent extensibility. According to SRVs definitions, \textit{GPTScan} can directly migrate by modifying natural language properties and scenario descriptions. Additionally, \textit{Slither} can be extended using \textit{SlithIR} intermediate languages within its modular detection framework.

\noindent \textbf{External Validity:} One external threat is the generality of vulnerability definitions. To ensure that SRVs reflect real-world issues, we collected and analyzed a total of 1,419 open-source security audit reports from 37 security teams. This is the first work to define and detect \textit{Signature Replay Vulnerabilities} in smart contracts. These types of vulnerabilities represent real issues encountered during development, and all identified vulnerabilities were sourced from authentic security reports.

% One external threat is the universality of vulnerability definitions. To ensure that the defined signature replay vulnerabilities reflect real-world issues, we collected and analyzed 1,419 public security audit reports from 81 smart contract audit and security teams. This is the first work to define and detect signature replay vulnerabilities in smart contracts, classifying them into five categories, all traceable to relevant security audit reports. Another external threat is that our detection results are based on publicly available datasets, which may not fully reflect the latest contracts. Additionally, the detection results focus on the behavior of signature reuse caused by the vulnerability, not on its exploitability, so it cannot be guaranteed that all detections are exploitable or profitable. We strive to ensure precision and accuracy in detecting signature replay behavior, but manual verification of exploitability or profitability can be difficult and subjective.

\vspace{-0.15cm}
\section{Related Work}
\subsection{Defining and Detecting Bugs in Contracts}
In recent years, the discovery and detection of vulnerabilities in smart contracts have become a key focus in blockchain security. As developers' awareness and mitigation technologies advance, common vulnerabilities like \textit{Integer Overflow}~\cite{overflow} and \textit{Unchecked Return Values}~\cite{uncheckreturn} have significantly decreased in real production environments. However, continuous innovation in DeFi applications has led to new vulnerabilities characterized by more complex states and divergent execution paths. To define and detect these vulnerabilities, existing work has summarized specific vulnerability types by investigating data from real production environments. For example, Chen et al.~\cite{chen2020defining} investigated \textit{Ethereum} \textit{StackExchange} posts and actual contracts, summarizing 20 common code defects and highlighting five high-risk protocol errors.  Zhang et al.~\cite{cryptoapi} studied \textit{Ethereum} transactions, contracts, and \textit{StackExchange} posts, identifying five major obstacles in developers' cryptographic tasks and providing a practical guide to improve the development experience. Yang et al.~\cite{nftguard} analyzed \textit{StackOverflow} posts, defined five common defects in NFT contracts, and proposed the \textit{NFTGuard} symbolic execution tool \textit{ NFTGuard } to automatically detect these defects.

% This paper is the first to define and detect contract \textit{Signature Replay Vulnerabilities}. We summarized and defined five types of vulnerabilities through empirical analysis of real-world audit reports and designed the \textit{LASiR} tool for automated detection.

\vspace{-0.3cm}
\subsection{LLMs for Smart Contract Security Research} 
With the development of LLM technology, its importance in smart contract security research has grown significantly. \textit{GPTScan} by Sun et al.~\cite{gptscan} uses the understanding of \textit{Generative Pre-training Transformer} (GPT)~\cite{chatgpt} to identify vulnerabilities by decomposing logical vulnerability types into scenarios and attributes, guiding GPT to match vulnerabilities. Wang et al.~\cite{tot}'s \textit{SmartInv} automatically detects "\textit{ machine unauditable bugs}" by reasoning about multimodal information, like source code and natural language, and uses the \textit{Tier of Thought} (ToT) prompting strategy to generate and analyze invariants to detect. Liu et al.~\cite{propertygpt} utilized state-of-the-art LLMs like \textit{GPT-4} to derive customized properties for unknown code from existing manual properties, such as audit reports. They created \textit{PropertyGPT} to verify the correctness of these properties.

% \textit{LASiR} leverages LLM's semantic understanding to assist in the static taint analysis of the signature state, guiding path reachability verification. By combining the strengths of LLM and program analysis, \textit{LASiR} ensures reliable analysis results.

\vspace{-0.3cm}
\section{Conclusion}
This paper presents the first empirical study on SRVs and defines five types. We designed \textit{LASiR}, which leverages LLMs to assist static taint analysis and integrates symbolic execution verification for efficient detection of SRVs. To evaluate detection performance, we collected 918,964 smart contracts on multiple blockchains. The results show that \$4.76 million in active assets are affected, with 19.63\% of \textit{Ethereum} contracts containing SRVs. Manual verification indicates \textit{LASiR} achieves a 95.83\% \textit{Recall}, with \textit{Precision} and \textit{F1-score} of 82.14\% and 88.46\%, respectively. LLMs significantly improve \textit{LASiR}’s performance, enhancing its detection capabilities.

\begin{acks}
This work was supported in part by the National Key Research and Development Program of China (No.~2023YFB2703600), the National Natural Science Foundation of China (Nos.~62276279, 62306344), the Guangdong Basic and Applied Basic Research Foundation (Nos.~2024B1515020032, 2024A1515010253), and the Major Key Project of Peng Cheng Laboratory under Grant PCL2025A07.
\end{acks}

% \nocite{*} %
\bibliographystyle{ACM-Reference-Format}
\bibliography{ref}

\end{document}